\UseRawInputEncoding
\documentclass[twocolumn]{aastex631}

\usepackage{soul}
\usepackage{natbib}
\usepackage{multirow}
\usepackage{graphicx}
\usepackage{tabularx}
\usepackage{verbatim}
\usepackage{color}
\usepackage{ulem}
\usepackage{subfigure}
\usepackage{amsmath}

\def\spose#1{\hbox to 0pt{#1\hss}}
\newcommand\lsim{\mathrel{\spose{\lower 3.0pt\hbox{$\mathchar"218$}}
     \raise 2.0pt\hbox{$\mathchar"13C$}}}
\newcommand\gsim{\mathrel{\spose{\lower 3.0pt\hbox{$\mathchar"218$}}
     \raise 2.0pt\hbox{$\mathchar"13E$}}}
\newcommand\rh{{\mathinner{R_\mathrm{h}}}}
\newcommand\rvir{{\mathinner{R_\mathrm{vir}}}}
\newcommand\rtwo{{\mathinner{R_\mathrm{200}}}}
\newcommand\mvir{{\mathinner{M_\mathrm{vir}}}}

\newcommand\Msun{{\mathinner{~ \mathrm{M_\odot}}}}

\newcommand\Mpc{{\mathinner{\mathrm{Mpc}}}}

\newcommand\kpc{{\mathinner{\mathrm{kpc}}}}

\newcommand\soutm{\bgroup\markoverwith
{\textcolor{black}{\rule[0.5ex]{2pt}{0.8pt}}}\ULon}

\submitjournal{ApJ}
\shorttitle{Dark matter traced by different components in and around galaxy clusters}
\shortauthors{Shin et al.}
\graphicspath{{./}{Figures/}}
\begin{document}

\title{
Spatial distribution of dark matter in and around galaxy clusters traced by galaxies, gas and intracluster stars in a simulated universe}

\correspondingauthor{Ho Seong Hwang}
\email{hhwang@astro.snu.ac.kr}

\author{Jihye Shin}
\affil{Korea Astronomy and Space Science Institute (KASI), 776 Daedeokdae-ro, Yuseong-gu, Daejeon 34055, Korea}
\author{Jong Chul Lee}
\affil{Korea Astronomy and Space Science Institute (KASI), 776 Daedeokdae-ro, Yuseong-gu, Daejeon 34055, Korea}
\author{Ho Seong Hwang}
\affil{Astronomy Program, Department of Physics and Astronomy, Seoul National University, 1 Gwanak-ro, Gwanak-gu, Seoul 08826, Korea}
\affil{SNU Astronomy Research Center, Seoul National University, 1 Gwanak-ro, Gwanak-gu, Seoul 08826, Korea}
\affil{Korea Astronomy and Space Science Institute (KASI), 776 Daedeokdae-ro, Yuseong-gu, Daejeon 34055, Korea}
\author{Hyunmi Song}
\affil{Department of Astronomy and Space Science, Chungnam National University, Daejeon 34134, Korea} 
\author{Jongwan Ko}
\affil{Korea Astronomy and Space Science Institute (KASI), 776 Daedeokdae-ro, Yuseong-gu,
Daejeon 34055, Korea}
\affil{University of Science and Technology (UST), Daejeon 34113, Korea}
\author{Rory Smith}
\affil{Departamento de Física, Universidad Técnica Federico Santa María, Avenida Vicuña Mackenna 3939, San Joaquín, Santiago, Chile}
\author{Jae-Woo Kim}
\affil{Korea Astronomy and Space Science Institute (KASI), 776 Daedeokdae-ro, Yuseong-gu, Daejeon 34055, Korea}
\author{Jaewon Yoo}
\affil{Korea Astronomy and Space Science Institute (KASI), 776 Daedeokdae-ro, Yuseong-gu, Daejeon 34055, Korea}
\affil{University of Science and Technology (UST), Daejeon 34113, Korea}

\begin{abstract}
To understand how well galaxies, gas and intracluster stars trace dark matter in and around galaxy clusters, we use the IllustrisTNG cosmological hydrodynamical simulation and compare the spatial distribution of dark matter with those of baryonic components in clusters. To quantify the global morphology of the density distribution of each component in clusters, we fit an ellipse to the density contour of each component and derive shape parameters at different radii. We find that ellipticity of dark matter is better correlated with that of galaxy mass-weighted number density, rather than with that of galaxy number density or galaxy velocity dispersion. We thus use the galaxy mass-weighted number density map as a representative of the galaxy maps. Among three different density maps from galaxies, gas, and intracluster stars, the ellipticity of dark matter is best reproduced by that of the galaxy map over the entire radii. The `virialized’ galaxy clusters show a better correlation of spatial distribution between dark matter and other components than the `unvirialized’ clusters, suggesting that it requires some time for each component to follow the spatial distribution of dark matter after merging events. Our results demonstrate that galaxies are still good tracers of dark matter distribution even in the non-linear regime corresponding to the scales in and around galaxy clusters, being consistent with the case where galaxies trace well the matter distribution in cosmologically large scales. 
\end{abstract}


\section{INTRODUCTION}

Galaxy clusters are important probes in 
  studying cosmology and structure formation \citep{all2011}. In particular, mapping the mass distribution in and around them provides valuable information that can be used for various comparisons between observations and simulations for testing structure formation models. The large masses
  and three baryonic ingredients\footnote{The total mass of galaxy clusters is $\gtrsim 10^{14}$ M$_\odot$; the contributions
 of dark matter, galaxies, intracluster objects and hot gas are $>80\%$, $1-2\%$, $<1\%$ and $5-15\%$, respectively \citep{boh2002,rud11}.
  (i.e. galaxies, intracluter objects, and 
  hot gas with temperature $T_{\rm gas}\gtrsim10^7$ K)
 of galaxy clusters provide various tools to trace the underlying dark matter distribution: e.g., galaxy redshift surveys for dynamical analysis \citep{gel2013,hwa2014,son19} optical/near-infrared imaging surveys for weak-lensing analysis \citep{oka2010,ume20} and X-ray and microwave surveys for the analysis of hot gas \citep{ett2013,Planck13}.}.

The total mass of clusters as a function of redshift
  can offer useful constraints on cosmological parameters
  \citep{vik2009,boc2019}.
The radial profile of cluster masses derived from observations 
  can not only be directly
  compared with the (universal) profile of dark matter halos in simulations
  (e.g., \citealt{bul2001}), but also be used
  for comparisons of mass measurements from different tracers
  \citep{gel2013,rin2013}.
  
The two-dimensional maps of cluster mass distribution also provide
  strong constraints on the mass growth of clusters and
  the formation of galaxies around them:
  e.g., existence of dark structures without galaxies \citep{clo2012,jee2014},
  and subhalo mass function \citep{oka2014}. 
However, such maps can highly depend on the systematics 
  of the tracers used for the map construction,
  so it is very important to cross check the mass measurements
  from different methods \citep{rin2016,sohn2018}.
In this regard, \citet{gel2014} showed a seminal example of such a 
 comparison between weak lensing and redshift surveys
  for Abell 383 (see their Figure 8).
Their results indicate that the weak-lensing cluster masses 
 could be affected by foreground/background structures even within the virial radii of clusters (see also \citealt{hoe01,dod04}).
This comparison demonstrates the importance of dense redshift surveys 
  to understand the mass distribution in weak-lensing maps.
\citet{hwa2014} extended such comparisons to nine relaxed clusters
  for more quantitative analysis using cross-correlation technique between the weak-lensing convergence map and the galaxy number density map.
This line of study was further extended to a merging cluster that has both weak lensing and redshift survey data exist \citep{liu2018}; many substructures with multiple merging processes in merging clusters make the comparison of mass distributions from different components more interesting as in the Bullet cluster \citep{clo04}. In these analyses, galaxies were used as test particles
  with the assumption that they trace dark matter well.
This issue has been tested since \citet{kai1984}, and
  the galaxies as biased tracers of dark matter
  is reviewed thoroughly in \citet{des1028}. 

Here we use the cosmological hydrodynamic simulations
  where we know well the distributions of dark matter and 
  other components including galaxies, gas and intracluster 
  stars in clusters.
Then the comparison of spatial distribution of dark matter with those of other components can provide an important test for the reliability of the use of galaxies to infer the dark matter distribution in galaxy clusters (i.e. non-linear regime).
Among many ways to analyze the spatial distribution of each component in clusters,
  we focus on the comparison of their global morphology; this includes ellipticity, position angle, and centroid offset of the projected density distribution of each component derived from the ellipse fitting method (see Section \ref{ellipse} for details).
We thus adopt the method fitting the contours 
  of mass or number density of each component
  with ellipse to derive shape parameters at different radii.
We note that the ellipse fitting may miss some possible small-scale
  differences among different components, 
  but we think that our approach can provide quantitative and 
  straightforward comparisons among different tracers; for example, see \citet{mon2019} for non-parametric methods. 

On the other hand, thanks to recent wide-field, deep images 
 available for galaxy clusters, the intracluster objects other than galaxies
 including globular clusters (e.g. \citealt{west1995,lee2010,ko2017}),
 planetary nebulae (e.g. \citealt{fel2004,lon2018}),
 and stars (e.g. \citealt{mih2005, ko2018})
 have become valuable probes of dark matter in clusters.
In particular, whether the intra-cluster light (ICL)
  could be an excellent tracers of underlying dark matter distribution or not
  has been seriously examined \citep{mon2019, alo2020, sam2020}
  in both observations and simulations.
A study focusing on the detailed connection 
  between ICL and dark matter using various methods
  will be a topic of separate papers in our group (Yoo et al., accepted).
Here we focus on the large-scale spatial distribution of galaxies
  along with their kinematics
  in and around clusters
  in comparison with other components including gas and intracluster stars.

Different mass accretion and/or merging histories of galaxy clusters can affect not only the current dynamical status of galaxy clusters, but also the spatial distribution of dark matter and other components therein \citep{moh93,buo95,ber00,lot04,par15,cui17,kim22}. To investigate how the different dynamical states are imprinted in the large-scale spatial distribution of each component around galaxy clusters, we divide the galaxy clusters into two sub-samples as `unvirialized’ and `virialized’, and compare the results between the two.

We describe the simulation data and the method for measuring the 
 shape parameters we adopt in Section \ref{data_method}, and
 compare one- and two-dimensional distributions of dark matter with those of galaxies (including two-dimensional number density and velocity dispersion maps), gas and intracluster stars in Section \ref{results}.
We discuss and summarize the results 
  in Sections \ref{discuss} and \ref{summarize}, respectively.
Throughout,
  we adopt cosmological parameters as in the TNG simulation: 
  $\Omega_{\Lambda,0}=0.6911$, $\Omega_{m,0}=0.3089$,
  $\Omega_{b,0}=0.0486$, $\sigma_8=0.8159$ 
  and $n_s=0.9667$ and $h = 0.6774$ \citep{planck16}.
All quoted errors in measured quantities are 1$\sigma$, and 
  spatial quantities and coordinates are expressed in comoving units.

\section{Data and Method}\label{data_method}

\begin{figure*}
    \centering
    \includegraphics[width=0.7\linewidth]{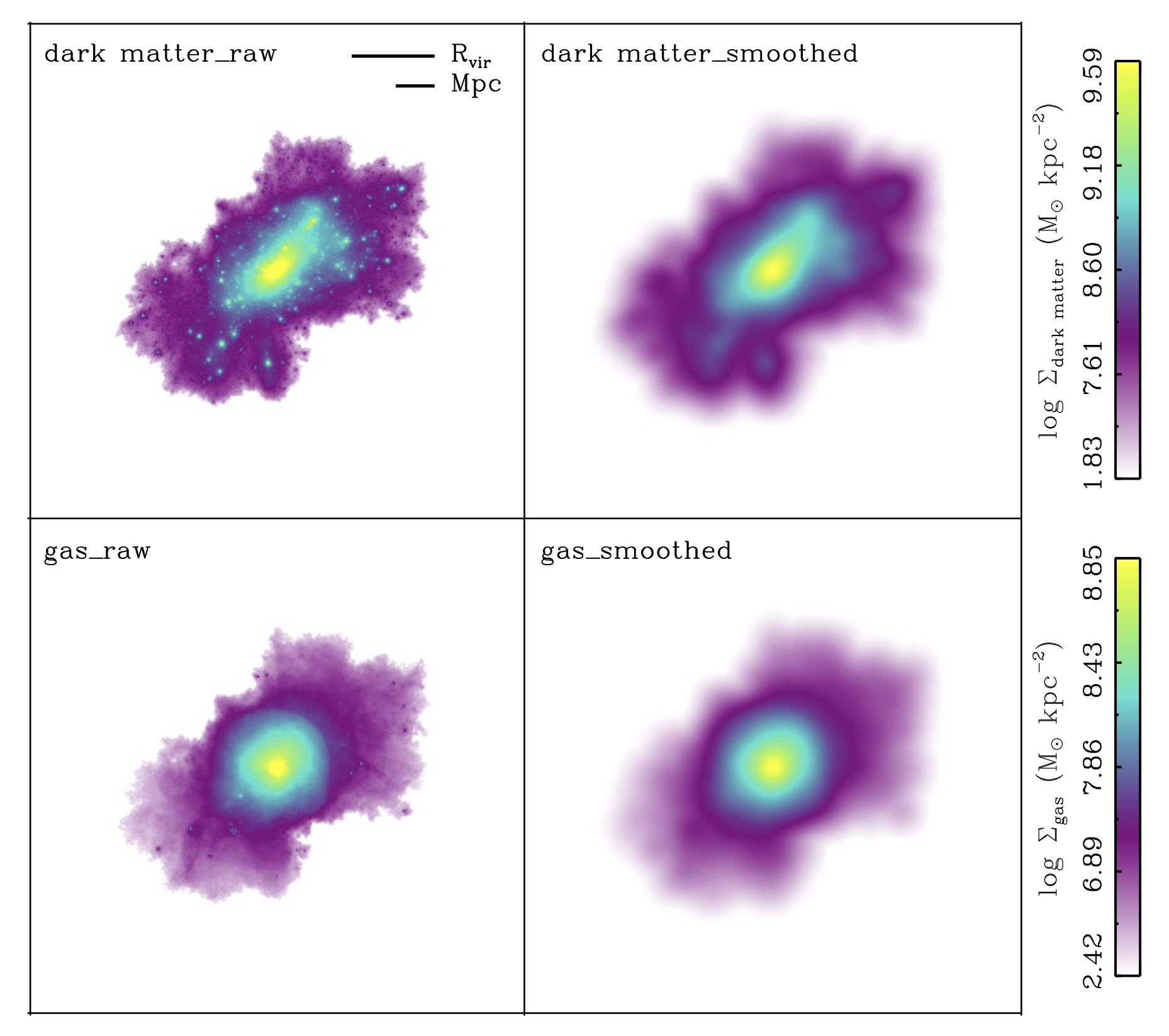} \\
    \caption{Dark matter (top) and gas (bottom) density maps in an example cluster. This cluster is the third most massive one in the TNG300. For its size information ($\rvir = 2.13~\Mpc$), the scale bars are presented in the upper-right corner of the top-left panel. All maps are xy projected. The original maps obtained by counting particles/cells in each pixel and the gaussian-smoothed, final maps are shown in the left and right panels, respectively. These maps are color-coded according to the mass density. For better visibility, the upper and lower limits are determined by the maximum pixel value in the final map and the minimum pixel value in the original map, respectively. The color scale is arbitrary.}
    \label{fig:making1a}
\end{figure*}

\begin{figure*}
    \centering
    \includegraphics[width=1.0\linewidth]{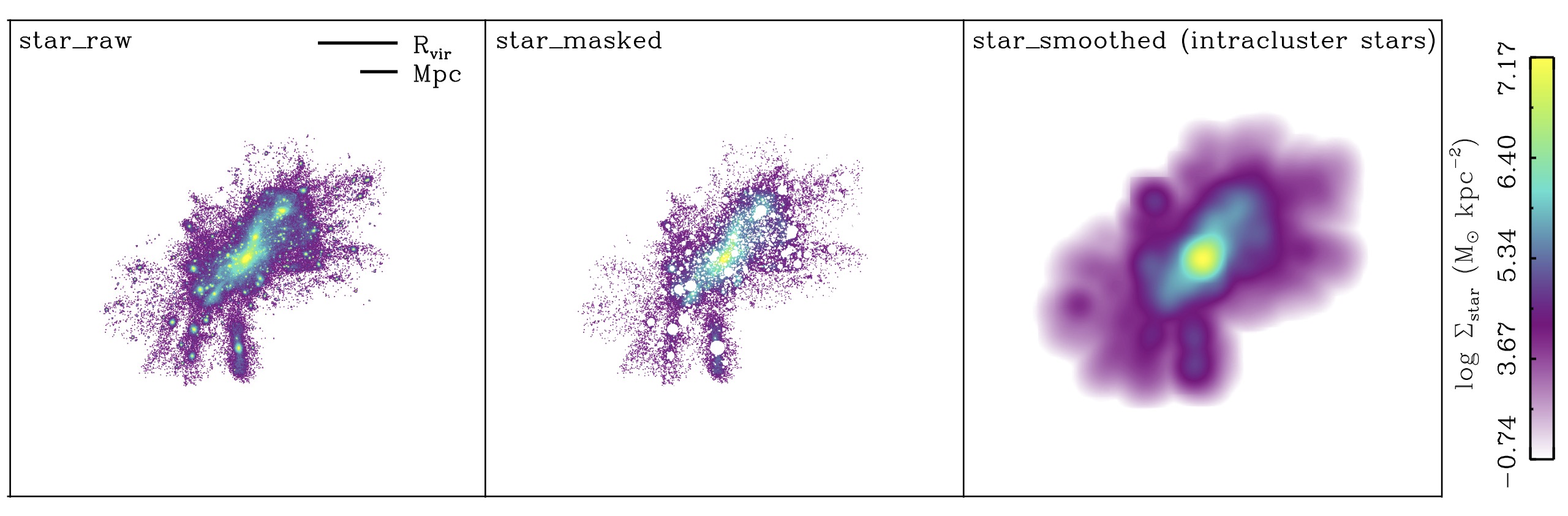} \\
    \caption{Similar to Figure \ref{fig:making1a}, but for the intracluster star component. The middle panel is added to show the intermediate process masking stars associated with member galaxies.}
    \label{fig:making1b}
\end{figure*}

\begin{figure*}
    \centering
    \includegraphics[width=0.6\linewidth]{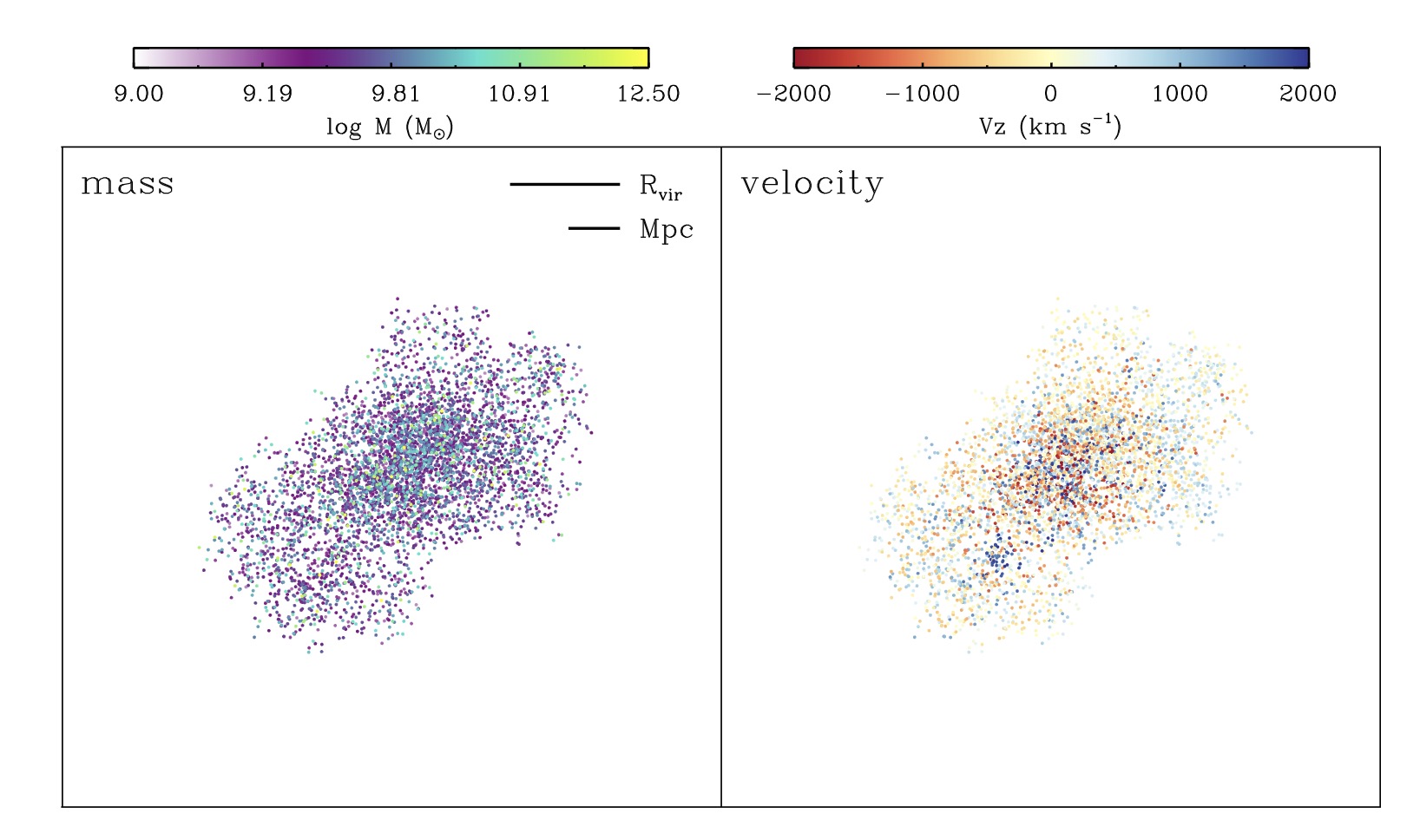} \\
    \caption{Spatial distribution of galaxies in the example cluster. This is color-coded by galaxy mass (left) and by velocity difference from the cluster mean motion (right). All of Figure 1-5 show the same object and are xy projected. Because the brightest cluster galaxy is extremely massive ($10^{15}\Msun$), the upper limit in the left color is determined by the mass of 2nd BCG.}
    \label{fig:making2a}
\end{figure*}

\begin{figure*}
    \centering
    \includegraphics[width=0.7\linewidth]{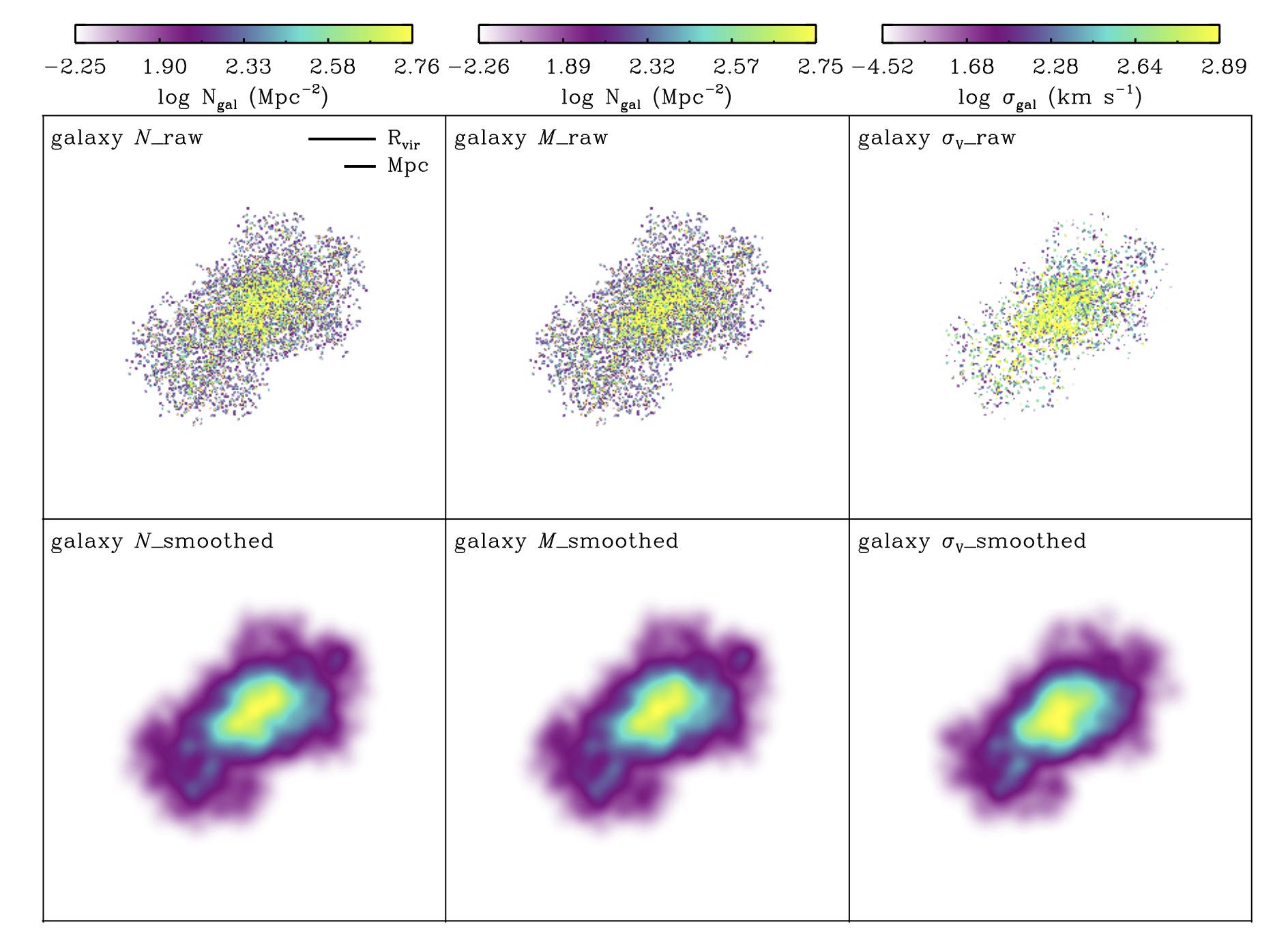} \\
    \caption{Similar to Figure~\ref{fig:making1a}, but for the galaxy maps. These are based on three different methods: number density (left), mass-weighted number density (middle) and velocity dispersion (right). The top panels are the original maps and the bottom panels are the smoothed ones with a Gaussian kernel, respectively.}
    \label{fig:making2b}
\end{figure*}

\begin{figure*}
    \centering
    \includegraphics[width=0.8\linewidth]{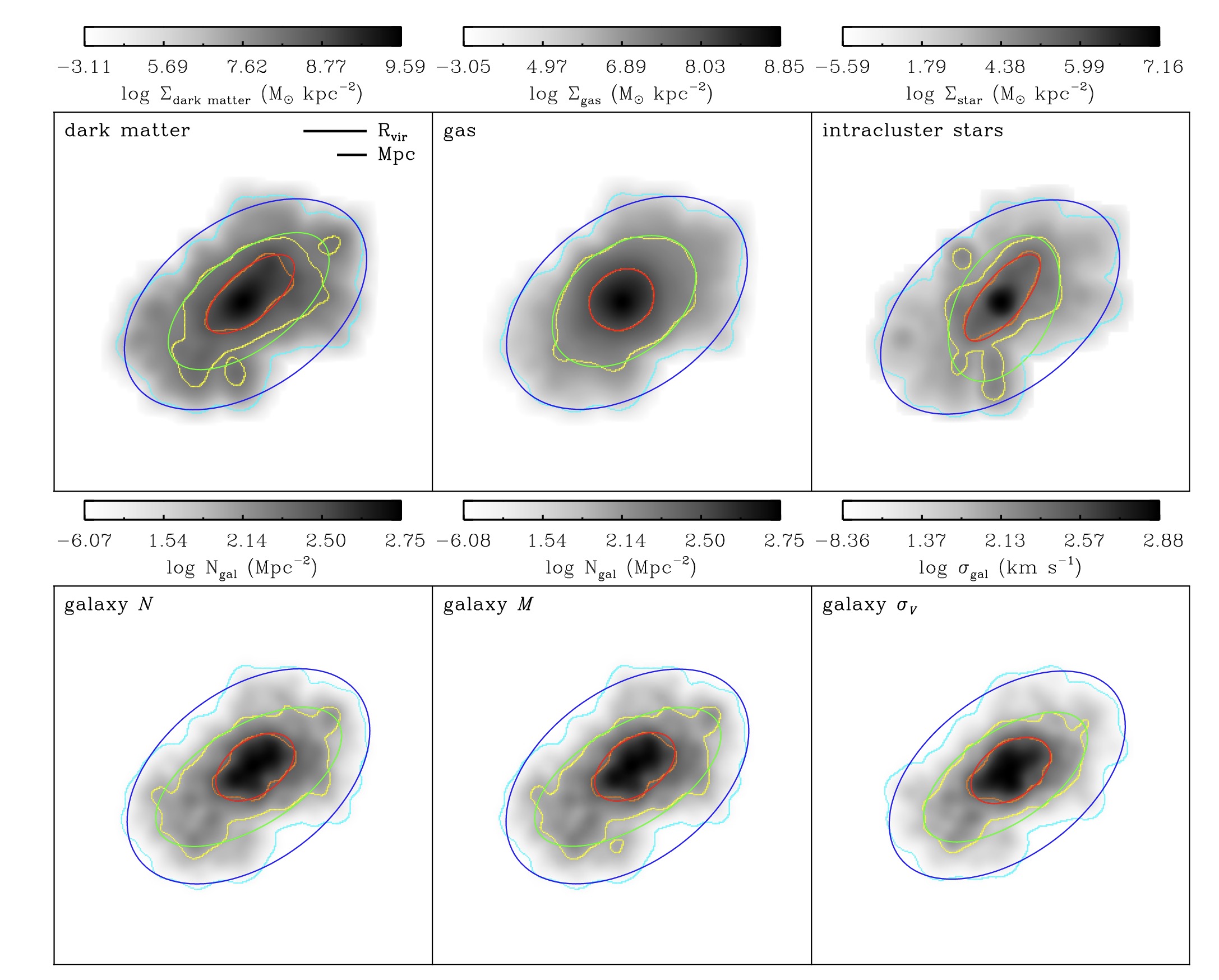} \\
    \caption{Results of the ellipse fitting for the xy projected maps of 
    the example cluster, which is the same as in previous Figures. 
    The background grey-scale maps for the upper panels show  
    dark matter (left), gas (middle), and intracluster stars (right), 
    while those for the lower panels are galaxy number density (left), 
    galaxy mass-weighted density (middle), and velocity dispersion (right),
    respectively. The orange, yellow and cyan contour lines represent 
    the inner, intermediate and outer shapes, respectively.
    The best-fitting ellipses, based on the contour lines, are 
    over-plotted in red, green and blue.}
    \label{fig:ellipse}
\end{figure*}

\subsection{Data}\label{data}
We use the publicly released data from the TNG300 \citep{nel19,pil18b,spr2018}, which is one of a suite of cosmological hydrodynamic simulations of the IllustrisTNG \citep{pil18a}: the successor of the Illustris simulation \citep{vog14}. The IllustrisTNG is composed of three different volumes whose one side length is roughly 300, 100 and 50 $\Mpc$: TNG300, TNG100, and TNG50, respectively. Among the TNG series, the TNG300 has the largest simulation box and thus enables us to obtain the most reliable statistics on galaxy clusters with the least cosmic variance even though the mass (spatial) resolution of dark matter particle is limited most as $5.9\times10^7\Msun$ (1.48 $\kpc$); the resolutions of dark matter for TNG100 and TNG50 are $7.5\times10^6\Msun$ (0.74 $\kpc$) and $4.5\times10^5\Msun$ (0.29 $\kpc$), respectively. Because we are interested in a comparison of the global morphology of the cluster’s matter distribution with that of each different component, rather than in the detailed structures that are highly affected by a choice of the resolution, we decide to use the TNG300 that gives best statistics for this study. Note that projected mass density maps generated for this study (see Section \ref{mapmaking}) are smoothed by a Gaussian kernel with a σ value of $90-240~\kpc$, and thus the $1.48~\kpc$ resolution of the TNG300 is much smaller than the resolution that we deal with in this study.

We use the properties of the FoF (friends-of-friends) halos listed in the group catalog, which are identified with the FoF algorithm with a linking length $b=0.2$ at $z=0$. To choose the galaxy cluster analogs from the catalog, we select massive FoF halos whose enclosed mass in a sphere of $\rvir$ or $\rtwo$ (the radius whose mean density is 200 times the mean density of the universe) is larger than $10^{14}\Msun$: the lower-mass limit typically used for galaxy clusters\footnote{The galaxy group analogs below the cluster mass range is behind of scope of this paper.}. The number of selected halos considered as galaxy clusters is 426 in total, whose mass and size is ranging $10^{14} < \mvir/\Msun < 2.9\times10^{15}$  and $0.9<\rvir/\Mpc<2.4$, respectively. We use this sample for following analyses. 
Here, we consider halos and subhalos in simulations as galaxy clusters and cluster galaxies in observations, respectively.

\subsection{Projected Mass Density Maps}\label{mapmaking}

Using the snapshot at $z=0$, we make projected mass density maps of dark matter, gas\footnote{We do not separate different stages of the gas (i.e. cold and hot). Most gas is dominated by hot one except for the very central region of individual galaxies, which is much smaller than typical resolution of the projected density map in this study (i.e. 30-80 $\kpc$).}, and star for each galaxy cluster. To do this, we first extract all particles/cells belonging to each cluster and then project them to three different directions of x-y, y-z, and x-z planes. The projected distribution of the particles/cells is counted in a 2-dimensional array of $900\times900$ pixels$^2$ using a count-in-cell (CIC) subroutine built in the IDL astronomy user's library (see the 1st columns of Figure~\ref{fig:making1a} and \ref{fig:making1b} ). Each dimension of the maps covers $\pm3\rvir$ from the cluster center, where the gravitational potential energy is at the minimum, and thus 15 pixels correspond to 0.1$\rvir$. 

Before making the smoothed image of the projected mass density, the local stellar density spikes originated from individual galaxies are removed because stars inside the galaxies tend to reflect the mass distribution of host galaxies rather than that of galaxy clusters; we are interested in global distribution of the diffuse intracluster stars that follows the cluster potential. Note that the individual galaxies are treated as a separated component. Using the subhalo catalog containing galaxy properties measured with the SUBFIND algorithm \citep{spr01}, the pixels out to 2$\rh$ from each galaxy center are masked, where $\rh$ is the radius containing half of each galaxy mass (see the 2nd column of Figure ~\ref{fig:making1b}). Very small-sized galaxies whose $2\rh$ is smaller than a half of pixel size are omitted from masking and treated as the diffuse intracluster stars. Note that we also apply no mask to the brightest galaxy in the cluster center (i.e. BCG) because the mask for the BCG can remove most of the central region of the cluster. 

We first use these projected mass density maps to calculate the one-dimensional density profile for each component (see Section \ref{1d}). We then smooth the projected mass density maps for the comparison of two-dimensional distributions (see Section \ref{2d}). To do that, we apply a Gaussian kernel with the standard deviation of 15 pixels to the maps using a GAUSS$\_$SMOOTH/IDL routine (see the last columns of Figure~\ref{fig:making1a} and \ref{fig:making1b}). The masked regions for the stellar mass density map are treated as non-existent pixels when pixel values of the unmasked regions are updated by smoothing, but their pixel values can be updated by those of the unmasked regions. Most of masked regions are smoothed out by nearby unmasked regions. However, a large mask (radius$>3\sigma$) produces some defects including image distortion because the routine is set to update a pixel value using the surrounding pixels inside the radius of $3\sigma$. Thus, if the mask size is too large (i.e. radius$>3\sigma$), it even leaves a hole in the image because all pixel values cannot be updated with a real number. We apply the empirical criteria as explained in Section {\ref{ellipse}} to reduce the large mask effect on the ellipse fitting results. Note that the smoothed stellar mass density map after masking represents the mass density of diffuse intracluster stars including the BCG. 

To examine in detail how galaxies can be good tracers of dark matter, we generate three different galaxy maps: number density, mass-weighted number density and velocity dispersion (see Figure~\ref{fig:making2a} and \ref{fig:making2b}). For this, we extract position, velocity, and stellar mass of the cluster member galaxies from the subhalo catalog. Here, the galaxy mass reaches down to $\sim1.2\times10^9\Msun$, which is the minimum mass (20 dark matter particles) to be identified as a galaxy with the SUBFIND algorithm. The galaxy number density, mass-weighted number density and velocity dispersion (galaxy \emph{N}, galaxy \emph{M}, and galaxy $\sigma_v$, hereafter) are calculated in a 2D array of $300\times300$ pixels$^2$, where each dimension covers $\pm3\rvir$ from the cluster center. Because the number of galaxies is much smaller than those of dark matter, gas, and star particles, we decrease the pixel number by three times compared to those of the other density maps with  $900\times900$ pixels$^2$ to enhance the statistical significance on galaxy counting. We then smoothed the 2D array with $\sigma=5$ pixels, which is comparable to that of density maps for dark matter, gas, and intracluster stars ($900\times900$ pixels$^2$ with $\sigma=15$ pixels). 
The member galaxies we use as test particles are distributed at discrete positions and have a wide mass range of $10^9-10^{12}\Msun$. Thus a simple linear mass-weight scheme in constructing galaxy density maps can leave many clumpy structures. To avoid this issue, we rescale the mass weight so that the member galaxies with minimum and maximum masses have the mass weights of one and 10, respectively (except the BCG). Here, galaxy $\sigma_v$ is a measure of velocity dispersion of galaxies within the pixel having at least three galaxies. To have a fair comparison of spatial distribution among different components and to allocate the same numerical error that arises from the ellipse fitting, we rescale the density maps of dark matter, gas and intra-cluster stars to be $300\times300$ pixels$^2$, which is the same as for the galaxy density maps.

\subsection{Ellipse Fitting}\label{ellipse}

To parameterize the spatial distribution of each component, we perform ellipse fitting on the six different projected maps of dark matter, gas, ICS, galaxy \emph{N}, galaxy \emph{M}, and galaxy $\sigma_v$ that are obtained in the previous section.
We apply the ellipse fitting to each contour determined at three different radii, $R/\rvir$ = 0.5, 1 and 2; these radii were chosen to study the structures of inner, intermediate and outer regions, respectively. We use the radii normalized by $\rvir$ to have a fair comparison among clusters.
 We first estimate the density level of the map at each radius
 from the average pixel value within the annulus with a width of 0.1$\rvir$.
We then extract the contour line at each level 
 with the {\small CONTOUR/IDL} routine
 and convert the contour line into pixel points 
 for the input of the ellipse fitting. 
If there are several peaks above the density level, the contour line may consist of several loops with different sizes (e.g. three yellow loops in the top-left panel of Figure \ref{fig:ellipse}). In that case, we only adopt the longest loop that best reflects the primary structure.
To prevent that the distortion caused by the large masks affects the fitting,
 the pixel points with the distance from large galaxies ($\rh$ $>$ 0.1$\rvir$)
 less than 2$\rh$ are excluded from the input 
 for the intracluster star density map.
We use the {\small MPFIT/IDL} package \citep{mar09} to derive the 
 shape parameters of the ellipse: centroid, major axis length, 
 ellipticity and position angle.
The best-fitting results of an example cluster 
 are shown in Figure~\ref{fig:ellipse}. 
 
Samples for the comparison are constrained to satisfy the following conditions. First, massive galaxies with $\rh>400\kpc$ should not be located within $2.5\rvir$ from the cluster center to avoid the effect of large masks on the ellipse fitting. There are 14 clusters excluded from the original sample of 426 clusters. They are mostly in a process of cluster (or group) merging and thus have a massive galaxy with $\rh>400\kpc$, which was the BCG (or BGG) of the recently merged cluster (or group). Second, the number fraction of the non-zero pixel used for estimating the level of the map at a given radius should be larger than 30\%. Last, the number fraction of the valid pixels along the contour line for a given level should be larger than 80\% to obtain reliable ellipse fitting results. Note that the second and third criteria are applied to individual density maps at different radii, while the first criterion to individual clusters.  A total number of reliable samples after these selection criteria decreases to 14,973 from 19,170 (426 clusters $\times$ 3 projections $\times$ 5 components $\times$ 3 radii). 

\begin{figure*}
    \centering
    \includegraphics[width=180mm]{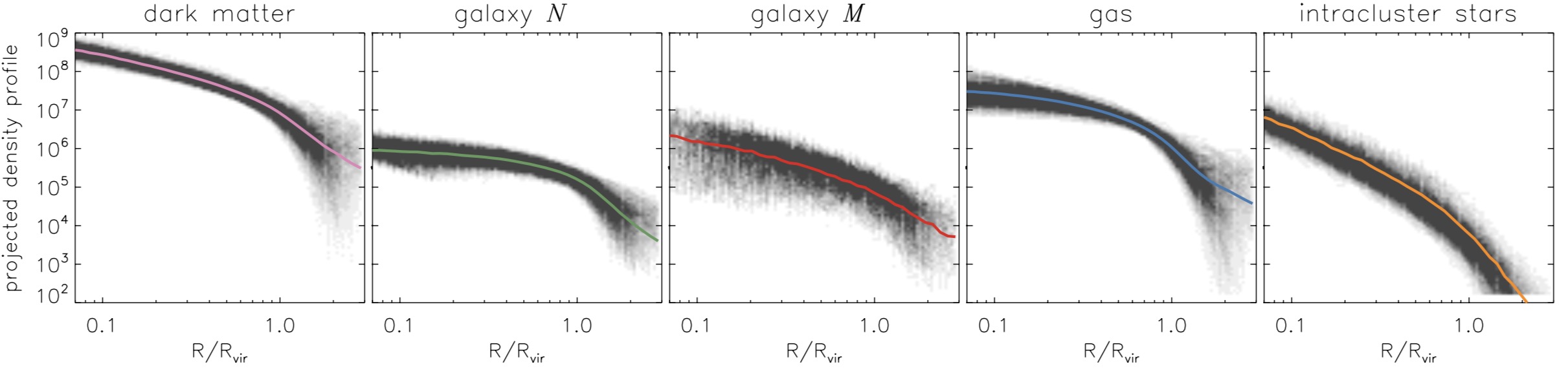} \\
    \caption{Projected density profiles of dark matter, galaxy \emph{N}, galaxy \emph{M}, gas, and intracluster stars. Individual density profiles of 426 clusters with three different projections are over-plotted with the gray lines, while their averaged profiles are shown together as the colored lines. The projected mass density profiles for dark matter, gas, and intracluster stars are in a unit of $\Msun\kpc^{-2}$. The projected galaxy number density profiles, galaxy \emph{N} and \emph{M}, are scaled up by a factor of $10^8$ to be shown in the same y-axis range of the other profiles.}
    \label{fig:prof_comp}
\end{figure*}

\begin{figure}
    \centering
    \includegraphics[width=80mm]{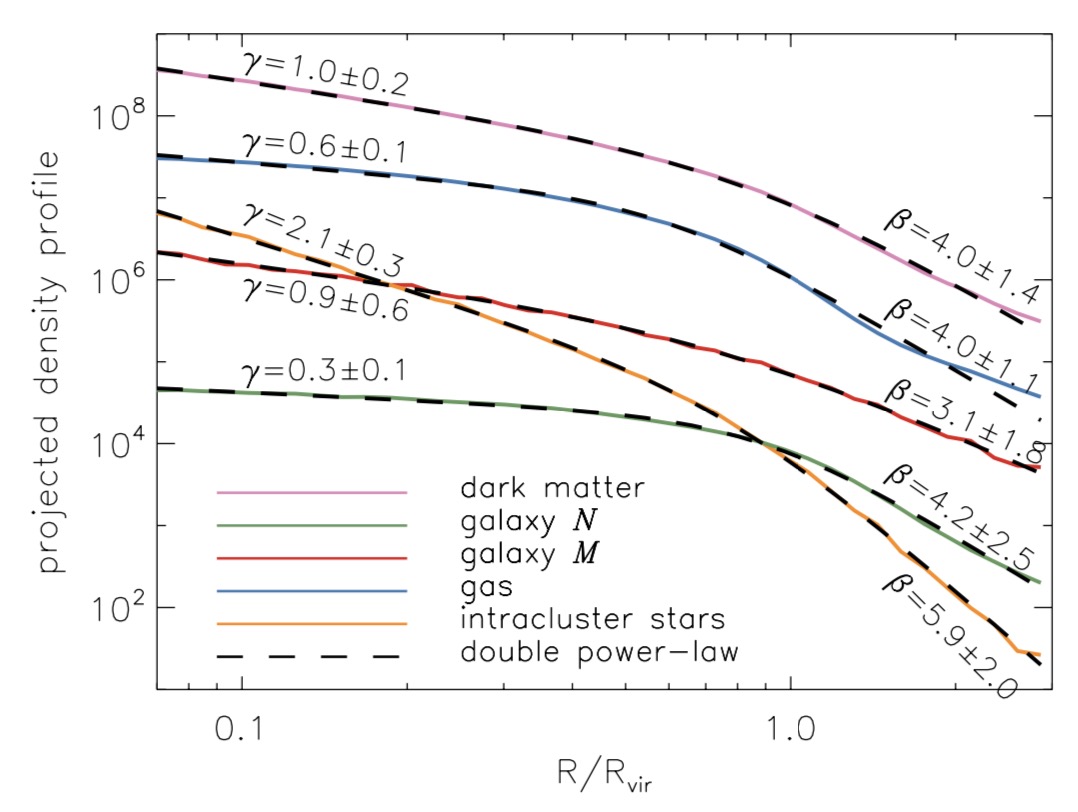} \\
    \caption{Averaged density profiles and their fitted lines with the double-power law. The overall density profiles are well described with the double-power law whose breaks are commonly in $0.7-1.0 \rvir$. Thus, the density profiles can be approximated as two separate power-laws with slopes of $\gamma$ and $\beta$ for $R/\rvir<1$ and $R/\rvir>1$, respectively. }
    \label{fig:prof_all}
\end{figure}
\section{RESULTS}\label{results}

\subsection{1D Radial profiles}\label{1d}

We first compare projected radial density profiles among different components. 
To calculate azimuthally averaged radial density profiles of dark matter, gas and intracluster stars, we use the projected mass density maps before smoothing: 
the 1st column of Figure \ref{fig:making1a} (dark matter, gas)  and the 2nd panel of Figure~\ref{fig:making1b} (intracluster stars). The galaxy number density profiles with and without mass-weight are calculated from the 1st row of Figure~\ref{fig:making2b}. 
Figure \ref{fig:prof_comp} shows the radial density profile of the 426 clusters 
for each component: dark matter, galaxy \emph{N} (i.e. number density without mass weight), galaxy \emph{M} (i.e. number density with mass weight), gas, 
and intracluster stars, as well as their averaged density profiles. Since the dark matter particle mass is fixed as $5.9\times10^7\Msun$, the dark matter mass density profile can be compared with the galaxy number density profiles 
even though their absolute value can not be directly compared to each other. 

\begin{table}[ht]
\caption{Parameter set for the double-power law function} 
\centering 
{\footnotesize
\begin{tabular}{r c c c c} 
\hline\hline 
Component & $\alpha$ & $\beta$ & $\gamma$ & $r_0$ [$\rvir$] \\ [0.5ex] 
\hline 
dark matter        & $2.0\pm3.2$ & $4.0\pm1.4$ & $1.0\pm0.2$ & $0.9\pm0.8$ \\ 
galaxy \emph{N}    & $3.4\pm1.5$ & $4.2\pm2.5$ & $0.3\pm0.1$ & $0.9\pm1.4$ \\
galaxy \emph{M}    & $1.9\pm2.3$ & $3.1\pm1.8$ & $0.9\pm0.6$ & $0.8\pm2.3$\\
gas                & $3.5\pm4.8$ & $4.0\pm1.1$ & $0.6\pm0.1$ & $0.7\pm0.3$\\
intracluster stars & $2.7\pm4.3$ & $5.9\pm2.0$ & $2.1\pm0.3$ & $0.8\pm0.8$\\ [1ex]
\hline 
\end{tabular}
}
\label{table} 
\end{table}

We fit the averaged density profiles with a double-power law of 
\begin{equation}
    \rho(r)=\frac{\rho_0}{(r/r_0)^{\gamma}[1+(r/r_0)^{\alpha}]^{(\beta-\gamma)/\alpha}},
\end{equation}
where $r_0$ is a scale radius, $\rho_0$ is a central density. Depending on $r/r_0$, $\rho$ can be approximated as follow: 
\begin{equation}
\begin{split}
 \rho \propto r^{-\gamma} & \quad \textrm{for} \quad r/r_0<<1,\\
 \rho \propto r^{-\beta}  & \quad \textrm{for} \quad r/r_0>>1,
\end{split}
\end{equation}
where $\alpha$ controls the sharpness of the break between the $\beta$ and $\gamma$ regimes. The double power-law function has been frequently used for describing the density profile of the dark matter structures formed in a cosmological context \citep{her90,zha96,nfw96}. It should be noted that the density profiles in Figure \ref{fig:prof_comp} are compiled from the projected density maps, so our fitting results with the double power-law function should be compared with those in other studies based on similar projections (e.g. \citealt{hol15, ume17,ko2018}).

Figure ~\ref{fig:prof_all} shows the averaged density profiles and their fitting results, which are summarized in Table~\ref{table}. We adopt the standard deviation of the 426 profiles for each component as 1-sigma uncertainty for the Levenberg-Marquardt least-square fitting process using MPFITFUN/IDL routine \citep{mar09}. Since the fitted profiles for the individual components have similar breaks around $r_0 = 0.7-1.0 \rvir$, they can be approximated by a broken power-law bent at similar $r_0$. Interestingly, the Galaxy \emph{M} profile shows the least conspicuous break. It’s inferred that is from different radial profiles of the high- and the low-mass galaxies inside the cluster region. Because the higher (or lower) mass-weight is more frequently given for galaxies locating at the center (outskirt) region, the slope for the Galaxy $M$ profile for $R/\rvir<1$ can be enhanced compared to that of the Galaxy $N$.  In the inner region of $R/\rvir<1$, the dark matter density profile can be approximately described as a power-law of $\gamma=1.0\pm0.2$; the galaxy \emph{M} gives $\gamma=0.9\pm0.6$, the most similar to that of dark matter. The dark matter profile in the outer region of $R/\rvir>1$ can be described as a power-law with $\beta = 4.0 \pm 1.4$ despite the scatters in the outskirts of the halo. The $\beta$ for the dark matter profile shows no significant difference from that of the galaxy \emph{M}, $\beta = 3.1 \pm 1.8$. The gas and the galaxy \emph{N} components, respectively, give $\beta$ of $4.0 \pm 1.1$ and $4.2 \pm 2.5$, the most similar to that of the dark matter. Similar to dark matter, the gas shows a significant departure from the fit at the end of the profile. Considering the fitted slopes and their uncertainties, the galaxy \emph{M} profile described by a double-power law appears most consistent with that of dark matter over the entire radial range. The profile for Galaxy \emph{M} shows the least conspicuous break. We do not have comprehensive explanation for this, but can at least comment on the comparison of the profile between galaxy \emph{M} and \emph{N}. The inner slope of the profile for galaxy \emph{M} (i.e. $\gamma$) is slightly larger than that for galaxy \emph{N}, which may be because of the amplified mass weight for the BCG and/or of the high-mass galaxies segregated into the inner region. On the other hand, the difference in the outer slope of the two profiles is not statistically significant given the error of each profile.
\begin{figure*}
    \centering
    \includegraphics[width=180mm]{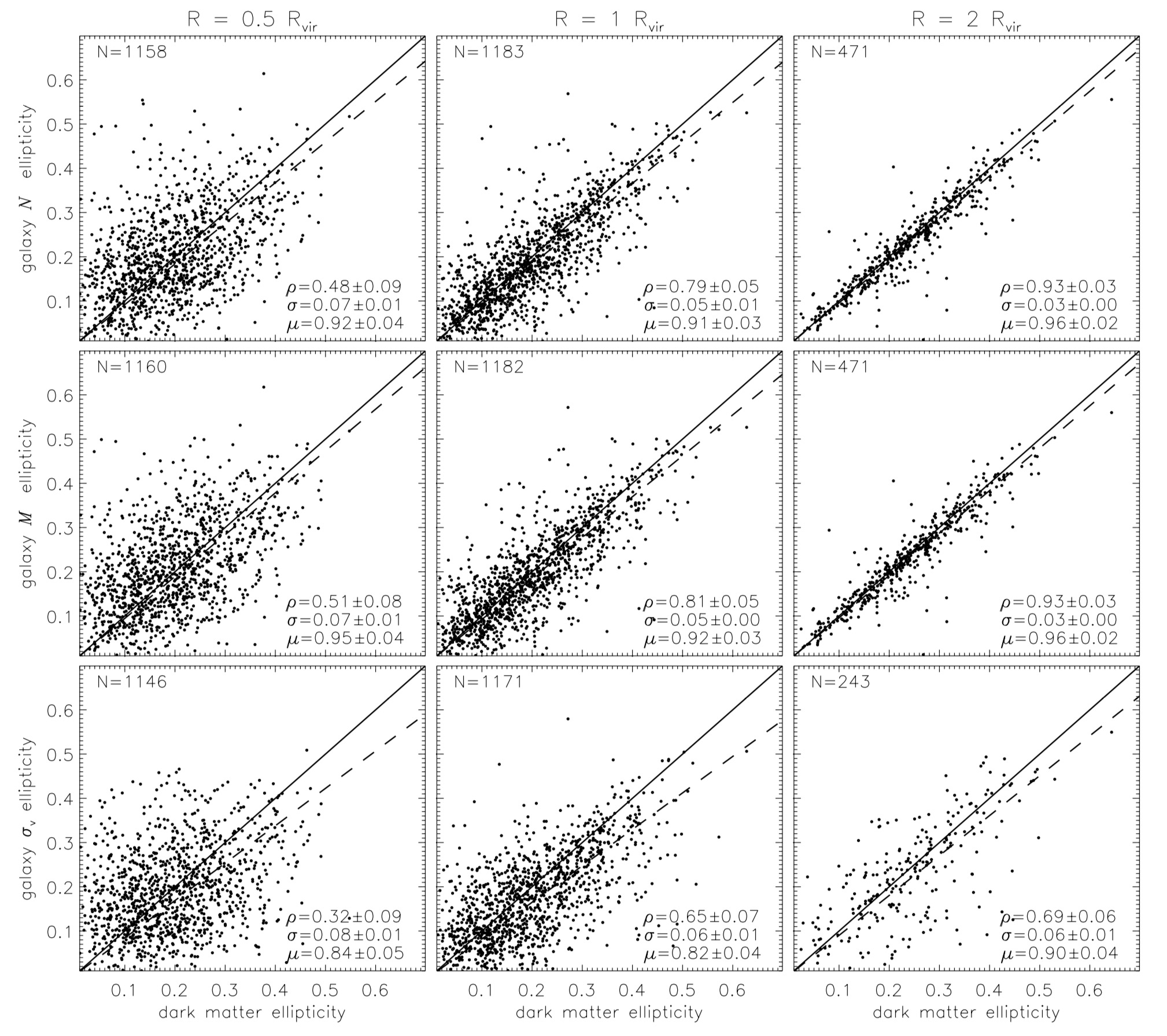} \\
    \caption{Comparison of ellipticity measured from the galaxy density maps (galaxy \emph{N}, galaxy \emph{M}, and galaxy $\sigma_v$) with that of dark matter at three different radii of 0.5, 1 and $2\rvir$. Correlation for each panel is measured by the Pearson correlation coefficient $\rho$, standard deviation from one-to-one relation $\sigma$, and fitted linear slope $\mu$. A dashed line indicates the best-fit relation. Number of samples with the reliable ellipticities is shown in each panel. See Section \ref{2d} for more details.}
     \label{fig:scatter_gal}
\end{figure*}

\begin{figure*}
    \centering
    \includegraphics[width=180mm]{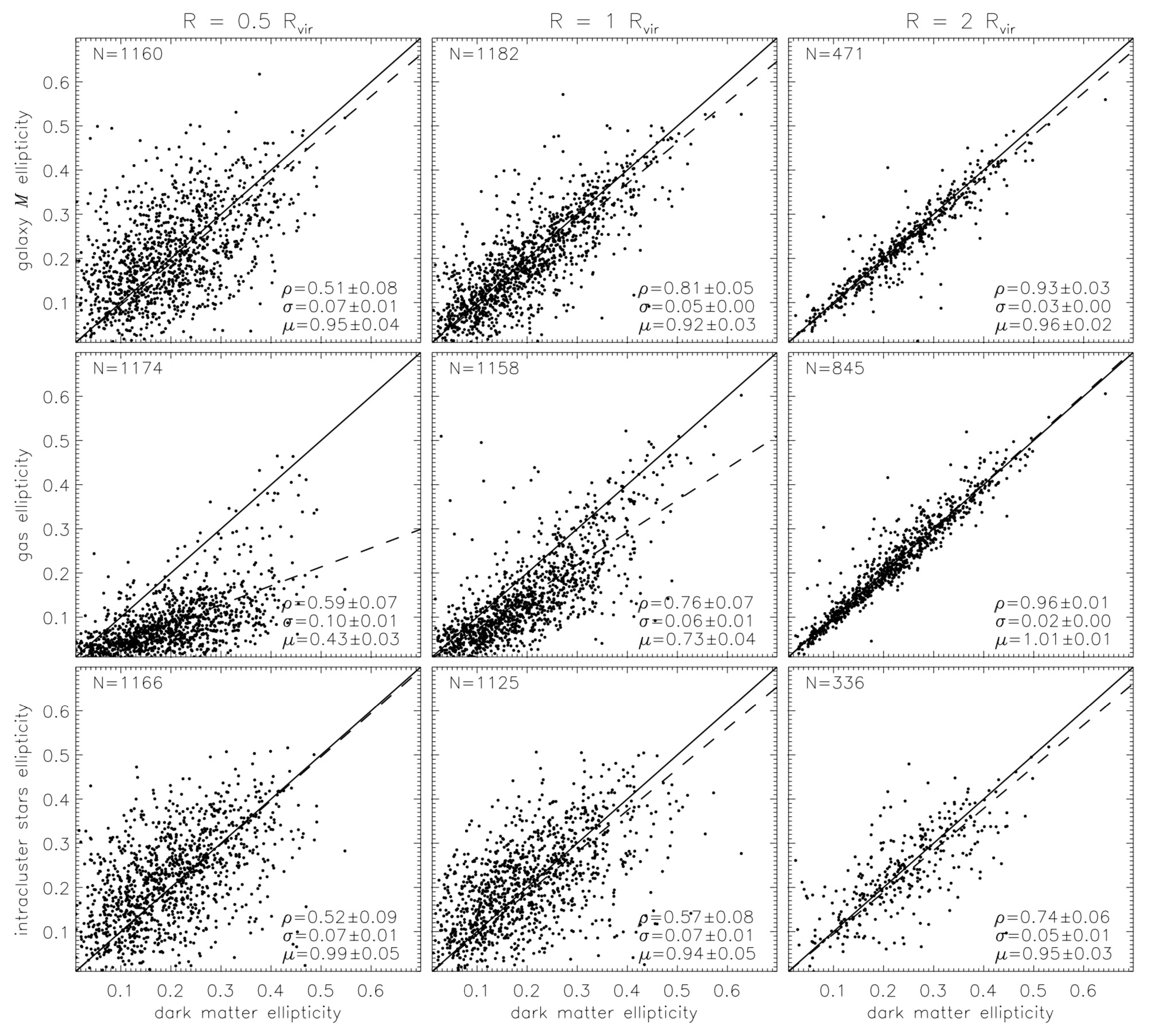} \\
    \caption{Similar to Figure~\ref{fig:scatter_gal}, but for galaxy \emph{M}, gas, and intracluster stars.}
     \label{fig:scatter_all}
\end{figure*}

\begin{figure*}
    \centering
    \includegraphics[width=180mm]{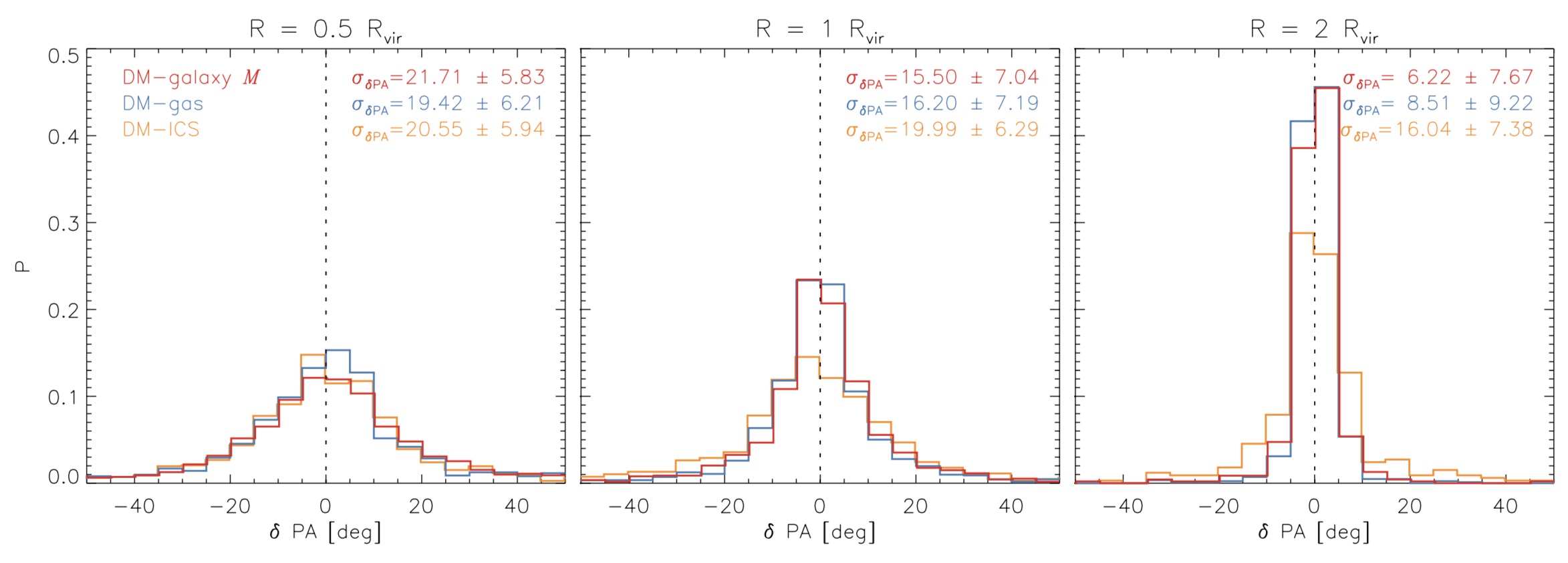} \\
    \caption{Histogram of position angle difference of each component with that of dark matter (shortly DM) at three different radii of 0.5, 1, and $2\rvir$. Dispersion from $\delta\mathrm{PA}=0$ is measured by the standard deviation $\sigma_{\delta\mathrm{PA}}$ for each component and radii. Intracluster stars are notated as ICS.}
     \label{fig:hist_pa_all}
\end{figure*}

\begin{figure*}
    \centering
    \includegraphics[width=180mm]{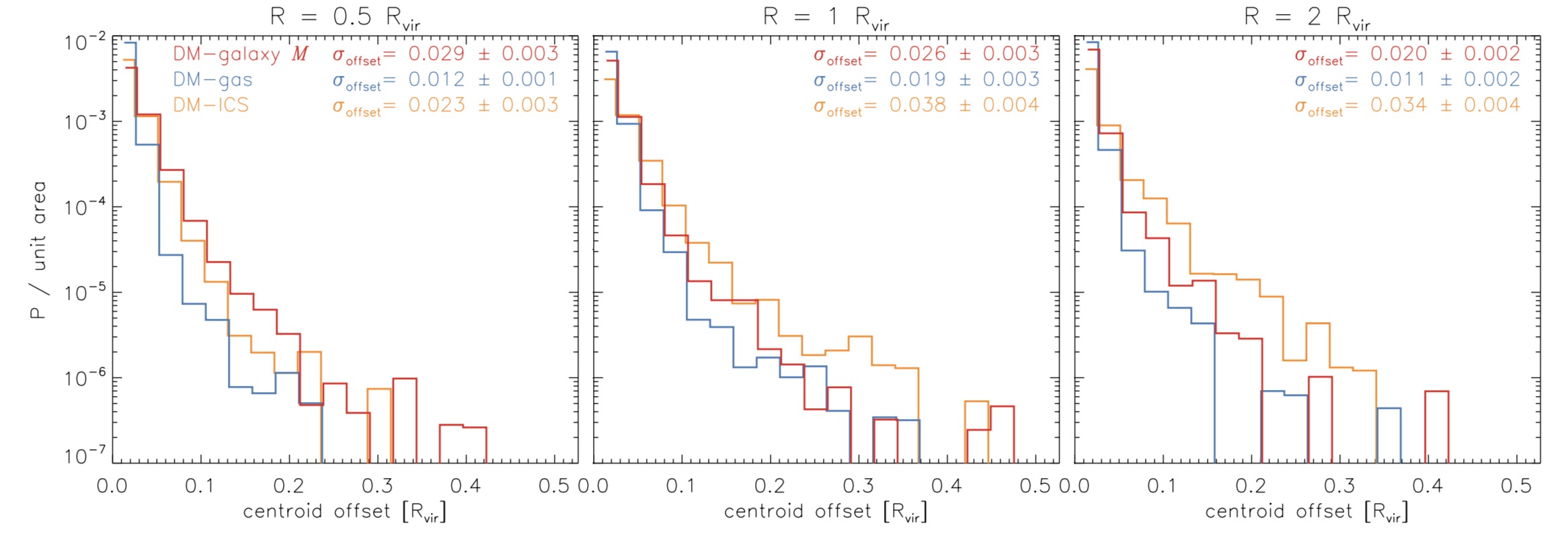} \\
    \caption{Similar to Figure~\ref{fig:hist_pa_all}, but for centroid offset of each component from that of dark matter. Histograms are normalized by annulus area, which is a function of radius centered at dark matter's centroid.}
     \label{fig:hist_offset_all}
\end{figure*}

\subsection{2D Spatial Distributions}\label{2d}

We compare the global morphology of the projected density distributions using the ellipse fitting results of ellipticity, position angle, and centroid offset. Before comparing the ellipse fitting results of three different components (i.e. galaxies, gas, and intracluster stars) with that of the dark matter, we first choose the representative galaxy map from galaxy \emph{N}, galaxy \emph{M}, and galaxy $\sigma_v$, whose ellipticity is best correlated with that of dark matter.

\subsubsection{Galaxy Components}\label{2d_galaxy}

Figure \ref{fig:scatter_gal} shows the correlations of the measured ellipticities of the maps between dark matter and galaxies (i.e. galaxy \emph{N}, the galaxy \emph{M}, and the galaxy $\sigma_v$) at three different radii of 0.5, 1, and $2\rvir$. We estimate the significance of the correlation using the Pearson correlation coefficient $\rho$, standard deviation from one-to-one correlation $\sigma$, and fitted slope $\mu$, which are shown in each panel. The ellipticity of each galaxy map at $R=2\rvir$ shows the strongest correlation with that of dark matter with the highest $\rho$ and the lowest $\sigma$ values. Particularly, the ellipticities of the galaxy \emph{N} and the galaxy \emph{M} show strong correlations with that of dark matter with $\rho=0.92\pm0.03$ and $0.93\pm0.03$, respectively. The correlation for the inner region of $R\leq1\rvir$ is getting weakened for all the cases not only by an intrinsic scatter coming from the different distribution between dark matter and galaxies (e.g. astrophysical effects including dynamical friction or baryonic feedback, \citealt{mas15,spr2018}), but also by a numerical error coming from the small number of pixels used for the ellipse fitting. The ellipticity measured by the galaxy \emph{N} and the galaxy \emph{M} still show strong correlations with that of dark matter even at $R=1\rvir$ with $\rho=0.79\pm0.05$ and $0.81\pm0.05$, while moderate correlations at $R=0.5\rvir$ as $\rho=0.49\pm0.09$ and $0.51\pm0.08$, respectively. In summary, the case of galaxy \emph{M} shows the highest $\rho$ values at all radii among the three galaxy maps. The galaxy $\sigma_v$ case differs from other cases significantly at $R=2\rvir$. Hereafter, we use the galaxy \emph{M} as a representative of the galaxy maps when we compare the spatial distribution of three different components of galaxies, gas, and intracluster stars with that of dark matter in and around galaxy clusters. 

\subsubsection{Comparison among Different Components}\label{2d_all}

The ellipticity of dark matter at three different radii of 0.5, 1, and $2\rvir$ is compared with those of the galaxies, gas, and the intracluster stars in Figure~\ref{fig:scatter_all}. At $R=2\rvir$, the ellipticity of the gas component shows the most prominent correlation with that of dark matter as $\rho=0.96\pm0.01$, which is significantly larger than that of intracluster stars ($\rho=0.74\pm0.06$) and slightly larger than that of galaxies ($\rho=0.93\pm0.03$). In the inner regions of $R\leq1\rvir$, the ellipticity measured for the gas component is getting biased to a lower value than that of dark matter due to `virialization’; kinetic energy of infalling gas is transformed to thermal energy (to be discussed in Section \ref{discuss}). Because gas distribution is getting circularized in the inner region, the fitted linear-slope for the ellipticity correlation is significantly reduced to $\mu=0.43\pm0.03$. Compared to the galaxy and the gas components, the ellipticity of the intracluster stars shows a weaker correlation with that of dark matter in general, except for the inner region. However, the difference of $\rho$ values among different components at the inner region is negligible, less than 1-sigma error. The radial difference of $\rho$ for the intracluster stars is relatively less prominent than those of other components. This seems to be because the BCG is not excluded, which makes a slightly better $\rho$ value in the inner region. In summary, among the maps based on different components, the ellipticity of the galaxy \emph{M} map shows the best correspondence to that of the dark matter map.

We now examine the difference of position angle (shortly, $\delta$PA) of the measured ellipse between each component and dark matter (see Figure~\ref{fig:hist_pa_all}). Because the PAs of different components at different radii do not show a systematic difference to that of dark matter, we show only the distribution of the scatter, so-called $\delta$PA, rather than the PA correlation of different components. Here, the $\delta\mathrm{PA}$ value closer to zero means that the ellipse's major axis of a component is more aligned to that of dark matter. The $\delta\mathrm{PA}$ distributions peak around zero ($-0.5^\circ<\overline{\delta\mathrm{PA}}<0.5^\circ$) with different standard deviation $\sigma_{\delta\mathrm{PA}}$ for different components and radii. The position angles of all the components show better alignments with those of dark matter at the outer region where the ellipse fit is less affected by numerical errors arising from the ellipse fitting. At $R=2\rvir$, galaxies show the smallest  $\sigma_{\delta\mathrm{PA}}$, but the difference from the other components is not so significant. We have perform the Kolmogorov-Smirnov test (K-S test) to examine how different the $\delta$PA distributions of galaxy $M$, gas, and ICS at $R=2\rvir$ are. The K-S test for the $\delta$PA distribution of galaxy $M$ and gas gives p-value of 0.45, indicating the null hypothesis that the two distributions are drawn from the same population cannot be rejected. The p-values from the K-S test between galaxy $M$ and ICs, and between gas and ICS are both $<0.001$, suggesting a significant difference between the two samples.

Figure \ref{fig:hist_offset_all} shows how the centroid of the ellipse measured for each component coincides with that of dark matter, which is measured by an offset of the centroid for each component from that of dark matter. Because the area of the annulus centered by the centroid of dark matter varies with radius, the count for each annulus changes with radius. Thus, to remove the difference introduced by this in the centroid-offset histogram, we normalize them by the area of the annulus. The normalized centroid offset distribution is highly peaked around zero, and the probability decreases exponentially with the centroid offset. The deviation of the normalized histogram from zero ($\sigma_{\mathrm{offset}}$) is measured to quantify the centroid consistency between each component and dark matter. The $\sigma_{\mathrm{offset}}$ for each component shows an insignificant change and/or trend for different radii. For a given radius, the gas component shows the lowest $\sigma_{\mathrm{offset}}$, which is different from other components by 2-sigma at least (to be discussed in Section \ref{discuss}). 

\subsection{Dependence on Dynamical Status of Galaxy Clusters}\label{virial}

\begin{figure}
    \centering
    \includegraphics[width=80mm]{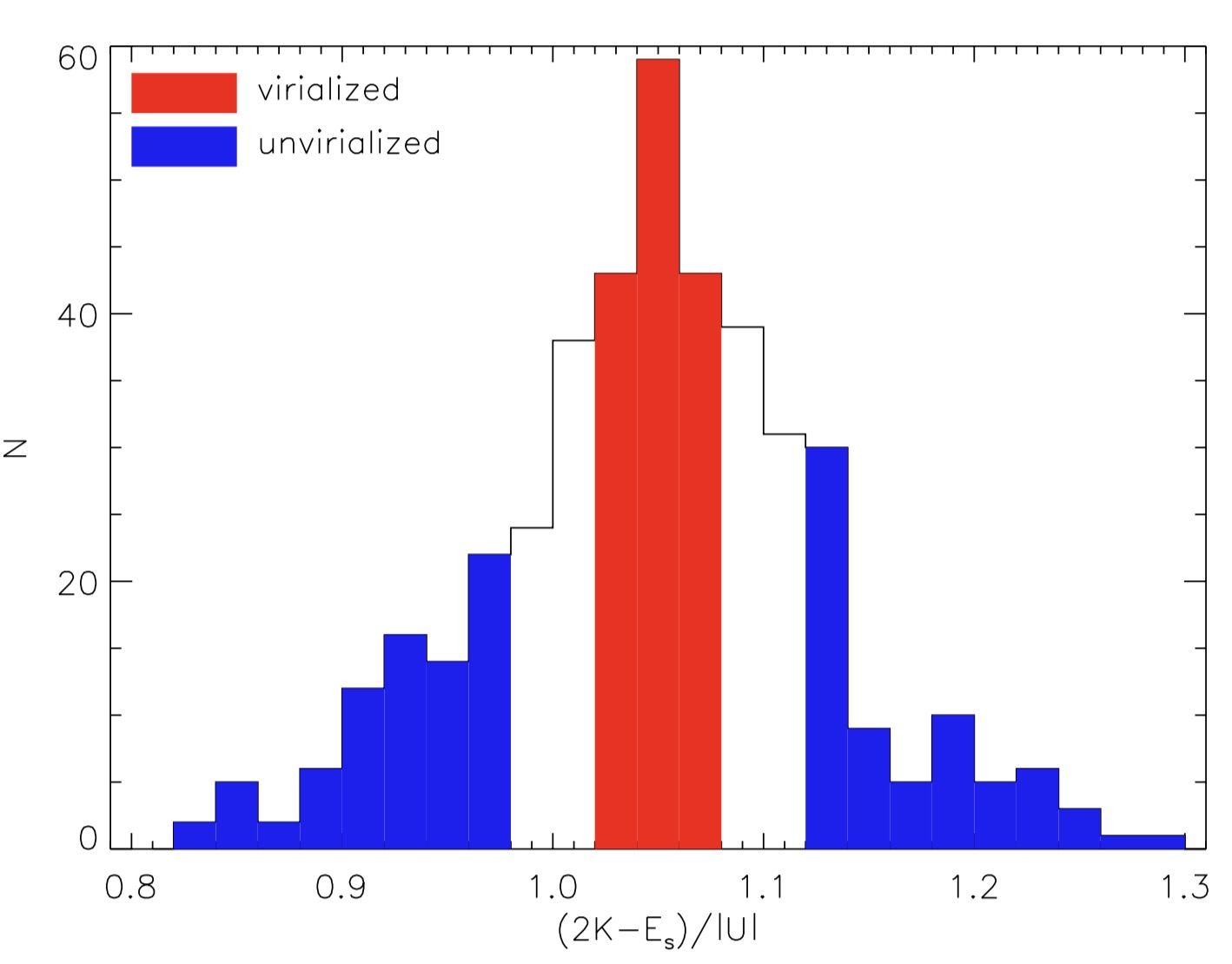} \\
    \caption{Histogram of virial ratio for the 426 clusters. Two sub-samples of `virialized' and `unvirialized' are selected to contain each the 200 closest and farthest clusters from the peak $Q$ value, which are filled with red and blue colors, respectively.}
     \label{fig:virial}
\end{figure}

To investigate how different dynamical states of galaxy clusters affect the correlation of global morphology between dark matter and other components (e.g. ellipticity), we divide galaxy clusters into two sub-samples using the virial ratio, which can be derived as 
\begin{equation}
    Q=\frac{2\mathrm{K}-\mathrm{E_s}}{|\mathrm{U}|},
\end{equation}
where $\mathrm{K}$ is a total kinetic energy, $\mathrm{U}$ is a total potential energy. Here, $\mathrm{E_s}$ is the energy from surface pressure at boundary, which is introduced to correct an assumption that the system is in complete isolation \citep{cha91}. Because a halo is embedded in a cosmological density field, particles close to the halo are continuously accreted across the halo boundary. To account for these particles giving pressure on the halo, the conventional virial theorem should be corrected by the additional term, so-called the surface pressure. We follow an approximated $\mathrm{E_s}$ formula of \citet{sha06} and \citet{cui17} for collisionless and collisional components, respectively. Figure \ref{fig:virial} shows the virial ratio distribution of 426 clusters. Despite the surface pressure correction, the $Q$ value peaks at 1.05, which is slightly higher than 1. 
Although we could not completely understand the cause of this difference, our $Q$ values are still good enough to construct subsamples for the comparison: i.e. virialized vs. unvirialized. We choose the 200 closest clusters to the peak $Q$ value and define them as the `virialized' sub-sample, while the 200 farthest clusters as the `unvirialized' sub-sample\footnote{We perform a similar analysis for two `unvirialized’ sub-samples:
86 clusters with `$Q<Q_{\mathrm{peak}}$' and 114 clusters with `$Q>Q_{\mathrm{peak}}$', where $Q_{\mathrm{peak}}$ is the peak of $Q$ distribution. Main results for the two `unvirialized’ sub-samples do not show any significant difference. A detailed discussion on the implication of the unvirialized clusters with different $Q$ values is beyond the scope of this paper.}.

\begin{figure*}
    \centering
    \includegraphics[width=180mm]{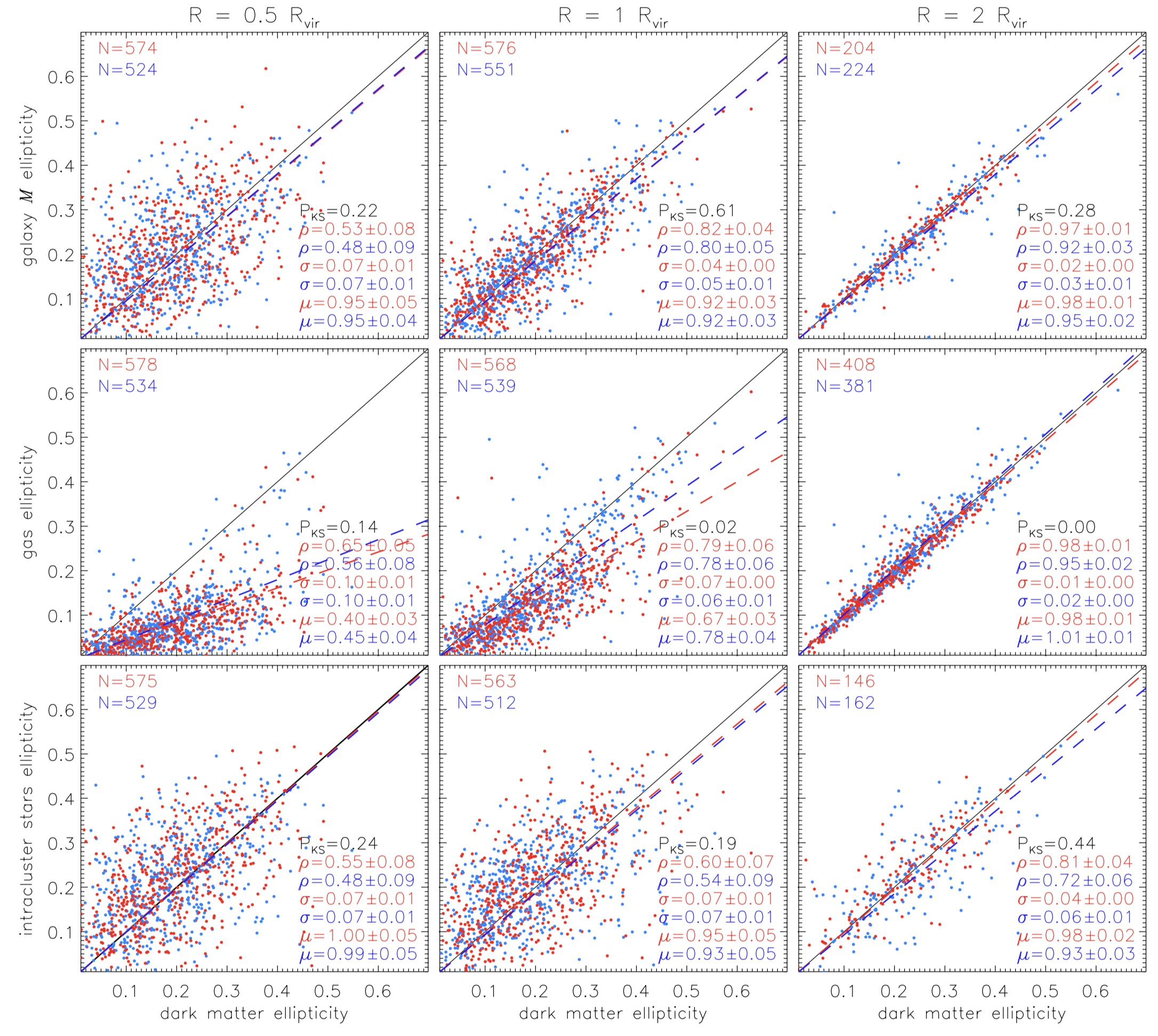} \\
    \caption{Similar to Figure~\ref{fig:scatter_gal}, but color-coded for two sub-samples of `virialized' (red) and `unvirialized' (blue). The p-value from the K-S test (P$_{\mathrm{KS}}$) is shown together for each panel.}
     \label{fig:scatter_sub}
\end{figure*}

Figure \ref{fig:scatter_sub} shows how the ellipticity correlation shown in Figure~\ref{fig:scatter_gal} is different for the two sub-samples. The `virialized' sub-sample shows a marginally higher $\rho$ value than the `unvirialized' sub-sample in all panels, but the difference is not so significant in most cases: only the outer region of $R=2\rvir$ shows a bit larger difference than the 1-sigma error. To statistically quantify the similarity of the ellipticity correlation between the `virialized’ and `unvirialized’ sub-samples, the two-dimensional KS-test is performed in each panel. The ellipticity distribution between gas and dark matter components at $R=1$ and $2\rvir$ shows a significantly low p-value as 0.02 and 0, respectively, which indicates the non-negligible difference between the two sub-samples. Note that the $\rho$ values can be used to compare how the ellipticity correlation from the two sub-samples are alike to each other, while the KS p-value is for the similarity of the ellipticity distributions between the two sub-samples, regardless of the ellipticity correlation.

\begin{figure*}
    \centering
    \includegraphics[width=180mm]{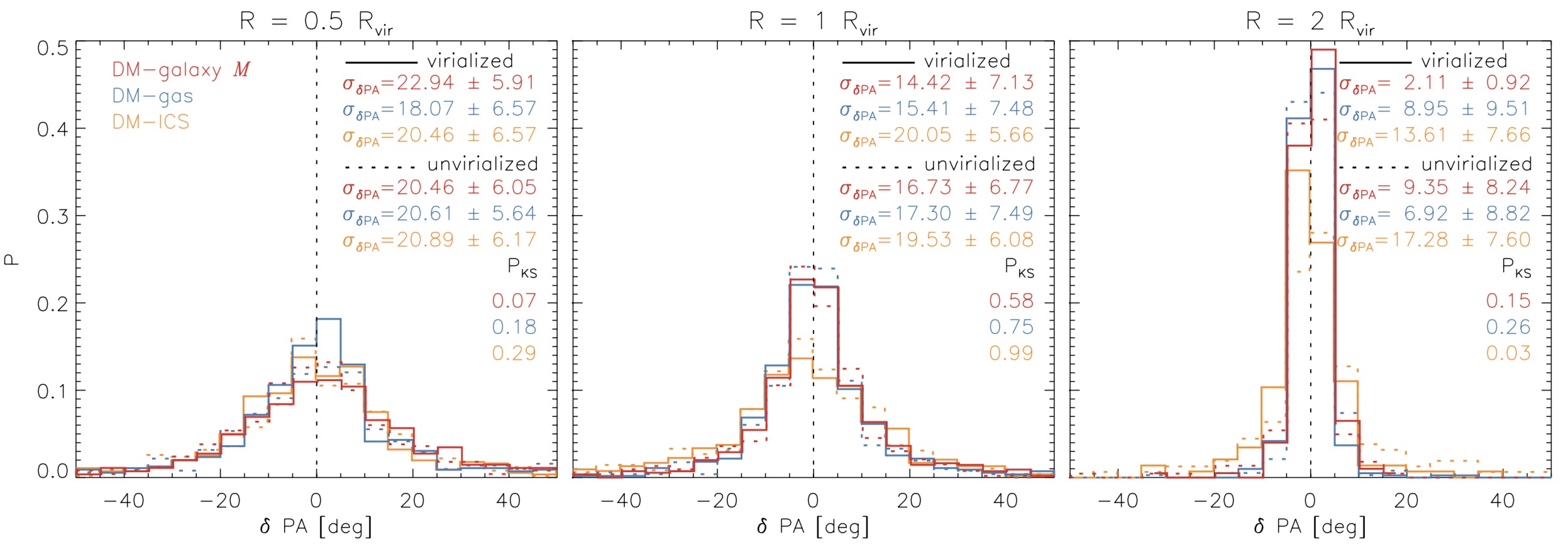} \\
    \caption{Similar to Figure~\ref{fig:hist_pa_all}, but for two sub-samples of `virialized' (solid line) and `unvirialized' (dashed line). The p-value from the K-S test (P$_{\mathrm{KS}}$) is shown together for each panel.}
     \label{fig:hist_sub_pa}
\end{figure*}

\begin{figure*}
    \centering
    \includegraphics[width=180mm]{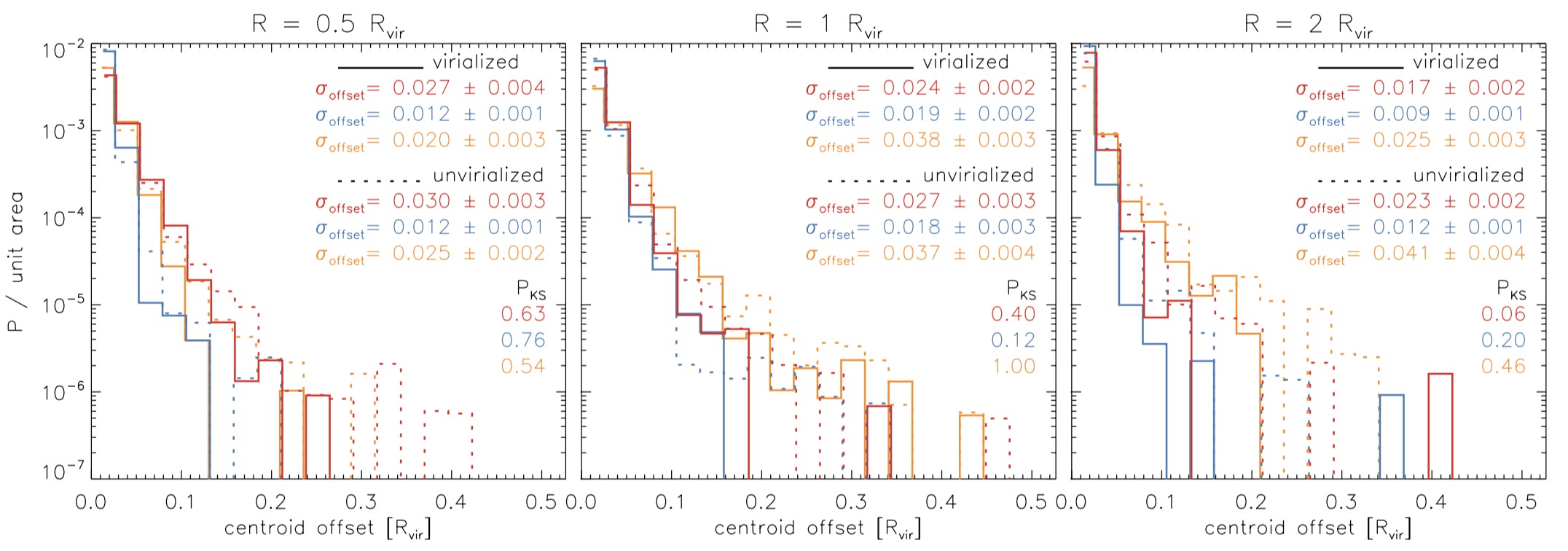} \\
    \caption{Similar to Figure~\ref{fig:hist_sub_pa}, but for centroid offset. }
     \label{fig:hist_sub_off}
\end{figure*}

Similar to the case of ellipticity, we examine how the two sub-samples show a difference for the $\delta$PA and the centroid offset distributions (see Figures \ref{fig:hist_sub_pa} and \ref{fig:hist_sub_off}). We find no statistically significant difference of the $\delta$PA distribution between the two sub-samples not only for all the components but also for all radii. In case of the centroid offset distribution, $\sigma_{\mathrm{offset}}$ of the `virialized' sub-sample at $R=2\rvir$ is significantly lower than that of the `unvirialized' sub-sample for all the components. It indicates that the centroid of ellipse measured at the outer region by each component is closer to that of dark matter for the `virialied' sub-sample, compared to the `unvirialized' one.

\section{DISCUSSION}\label{discuss}

It has been well known that galaxies are good tracers of matter field in cosmologically large scales, which is still in linear regime \citep{kai1984,des1028}. However, non-linear evolution in smaller scales can introduce a scale-dependent bias, which can not be easily predicted from a theory. Using the IllustrisTNG cosmological hydrodynamic simulations, \citet{spr2018} have studied the non-linear galaxy bias over a wide range of scales. In this study, we use the IllustrisTNG simulation and quantitatively compare the global spatial distribution of dark matter not only with that of the galaxies but also with those of gas and intracluster stars to better understand how well galaxies and other components (i.e. gas, intracluster stars) trace dark matter in and around clusters. 

We use the ellipse fitting method to reflect the global morphology of the galaxy clusters, and compare the ellipticity of the dark matter density map with that of three different galaxy maps of galaxy \emph{N}, galaxy \emph{M}, and galaxy $\sigma_v$, which have been widely used in observations to infer underlying matter distribution in cluster region \citep{hwa2014,zah2016,liu2018,sohn2020}. We find that the ellipticity of dark matter is better correlated with that of galaxy \emph{N} or \emph{M} rather than galaxy $\sigma_v$ over the entire radii. It is inferred that a small number of galaxies used for measuring $\sigma_v$ may not show the underlying dark matter distribution well. The measured ellipticity of the galaxy \emph{N} and \emph{M} and their correlation with that of dark matter are similar to each other (see Figure \ref{fig:ellipse} and \ref{fig:scatter_gal}). Between the galaxy \emph{N} and \emph{M}, the ellipticity measured from galaxy \emph{M} (mass-weighted galaxy number density) shows a slightly better correlation with that of dark matter even though the difference is not significant. This is consistent with the expectation from 
  the studies of large-scale structures focusing
  on the reconstruction of dark matter distribution
  with the halo/galaxy-mass based techniques 
  (e.g., \citealt{wang2009}, see also \citealt{hon21}
  for the importance of peculiar velocity information
  in the reconstruction).

We then use the similar ellipse fitting method to compare the global morphology of different components in galaxy clusters (i.e. galaxies, gas, intracluster stars). The three components show the strongest ellipticity correlation with dark matter at the outer region, and their correlations are getting weakened for the inner region. The larger scatter in the inner region seems to originate from the following two factors: 1) an intrinsic error coming from a fact that each test particle of galaxies, gas and intracluster stars tends to fail to trace the dark matter distribution at the inner region where the non-linear evolution takes place intensively (e.g. astrophysical effects including dynamical friction or baryonic feedback, \citealt{mas15,spr2018}), and 2) a numerical error originated from the smaller number of pixels for the inner contours, which results in the larger uncertainty on the ellipse fitting. To avoid the numerical error, a non-parametric comparison of contours without the fitting, such as the cross-correlation technique \citep{hwa2014} and the Modified Hausdorff distance method \citep{mon2019}, would be necessary in the future studies.  

In the outer region, the ellipticity of gas distribution shows the strongest correlation with that of dark matter even though the difference from the other components is not significant. The higher $\rho$ value of the gas component than that of the others could be because the gas component has a higher mass-resolution than other components so that the gas can describe the density distribution in a continuous way, similar to the case of dark matter. The gas ellipticity at the inner region, however, is getting biased to a lower value than that of dark matter, which is also shown as circular shape at Figure~\ref{fig:ellipse}. It is because infalling gas particles to the galaxy cluster are heated by a virial shock, and thus their kinetic energy is transformed to thermal energy: so-called `virialization'. The biased gas ellipticity shows the effect of the gas virialization only for the inner and middle regions (0.5 and 1 $\rvir$), while the effect has been believed to begin to occur at much further distance around 2-3 $\rvir$ \citep{zin18}. Although the ellipticity is biased to the lower value at the inner region, the gas component shows the smallest centroid offset with respect to dark matter's over the entire radii. It is inferred that gas particle can quickly adapt to the centrally concentrated mass distribution by collisions between themselves with much shorter timescale than those of other components. Separation between gas and dark matter of merging galaxy clusters is reproduced in the recent simulations of \citet{mou21}, which is similarly shown in Figure \ref{fig:hist_sub_pa}; the centroid of each component from the `unvirialized’ sub-sample is less aligned to that of dark matter. See the following paragraphs for details.

The diffuse intracluster light, which is not bounded to individual galaxies but to the cluster potential, is expected to trace the global matter distribution of clusters \citep{mon2019,alo2020,sam2020}. The intracluster stars in this study include the intracluster light and the BCG by excluding individual galaxies that are masked from the original stellar density map. However, the ellipticity of the intracluster star shows a weak correlation with that of dark matter compared to the other components in general. It is inferred that 1) the low-mass galaxies, not found with the SUBFIND algorithm ($< 1.2\times10^9 \Msun$) or whose $2\rh$ are smaller than a half of pixel size, are missing from the masking and thus lead to clumpy structures especially at the outer region where the low-mass galaxies are more abundant, and 2) the detailed tidal features of intracluster stars, which can trace interaction between galaxies, result in scatters on the ellipticity correlation since global spatial distribution is described by an ellipse only. 

Compared to the `unvirialized’ sub-sample, the `virialized’ galaxy clusters show a better correlation of spatial distribution between dark matter and other components at all radii even though the difference in the correlation between the sub-samples is not significant; the marginally higher $\rho$ value for the `virialized’ sub-sample can mean that each component follows better the spatial distribution of dark matter some time after merging events. The two-dimensional KS-tests for the ellipticity distributions of dark matter and gas show that there is a non-negligible difference between the `virialized’ and `unvirialized’ sub-samples only at $R=1$ and $2\rvir$. Since the `unvirialized' features can remain in the smaller-scale structures that may be missed by the the ellipse fitting, a different comparison focusing on detailed features would be necessary to better understand the difference between the sub-samples. 

Interestingly, the fitted linear-slope ($\mu$) of ellitpicity correlation between dark matter and gas of the `virialized' sub-sample at $R=0.5$ and $1\rvir$ shows a slightly lower value than that of the `unvirialized' sub-sample. It means that the gas ellipticity of the `virialized' galaxy clusters tends to be more biased to a lower value than the ellipticity of dark matter, compared to that of the `unvirialized' galaxy clusters. Because the `virialized' clusters tend to be growing in a continuous way by a smooth accretion, while the `unvirialized' clusters in a radical way by quite recent mergers, the smaller $\mu$ value for the `virialized' sub-sample can be circumstantial evidence that gas residing longer inside the galaxy clusters is heated more by a virial shock and thus circularized more.

Unlikely to the insignificant difference of the $\sigma_{\delta\mathrm{PA}}$ between `virialized’ and `unvirialized’ sub-samples for all components at all radii (see Figure \ref{fig:hist_sub_pa}), the $\sigma_{\mathrm{offset}}$ at $R=2\rvir$ shows a significant difference (see Figure \ref{fig:hist_sub_off}), where the centroid of each component from the ‘unvirialized’ sub-sample is less aligned to that of dark matter. This result is consistent with the observations showing that X-ray peak and the BCG tend to be separated more for non-cool-core clusters, which are believed to be associated with recent major merging events and thus tend to be unvirialized more \citep{kat03,san09,hud10,hof12,kim17}. At $R=0.5$ and $1\rvir$, on the other hand, the $\sigma_{\mathrm{offset}}$ between the two sub-samples does not show a significant difference. The non-parametric comparison (i.e. the Modified Hausdorff distance method, \citealt{mon2019}) should be applied to check whether the insignificant difference is originated from an intrinsic error coming from a fact that spatial distribution at the inner and middle regions is less sensitive to the recent merging events.

\section{SUMMARY}\label{summarize}

We use the IllustrisTNG cosmological hydrodynamical simulation to understand how well dark matter distribution in and around galaxy clusters is traced by galaxies, gas, and intracluster stars. Using the ellipse fitting method, we quantify the global morphology of dark matter distribution of galaxy clusters and compare the result (e.g. ellipticity) with those of galaxies, gas, and intracluster stars. Our primary results are as follows:\\  

(1) Among three different galaxy maps of galaxy \emph{N} (number density), galaxy \emph{M} (mass-weighted number density), and galaxy $\sigma_v$ (velocity dispersion), ellipticity of the galaxy \emph{M} shows the best correlation with that of the dark matter map over entire radii. \\

(2) The ellipticities of three different density maps from galaxies (galaxy \emph{M}), gas, and intracluster stars are compared with that of dark matter. We find that ellipticity of the dark matter map is best reproduced by that of the galaxy map.\\ 

(3) The ellipticity of the gas component shows the most prominent correlation with that of dark matter at the outskirt of galaxy clusters ($R=2\rvir$). However, in the inner region of $R\leq1\rvir$, it is significantly biased to a lower value than that of dark matter due to `virialization’. \\

(4) We have examined how position angle and centroid of the measured ellipse for each component coincides with that of dark matter: $\delta$PA and centroid offset, respectively. We find that $\delta$PA and centroid offset distributions do not show an insignificant difference and/or trend for different components and radii.\\

(5) The `virialized’ galaxy clusters show a better correlation of spatial distribution between dark matter and other components than the `unvirialized’ clusters. This can suggest that it requires some time for each component to follow the spatial distribution of dark matter after merging events.\\

This study demonstrates that galaxies are still good tracer of dark matter distribution in and around galaxy clusters, similar to the case where galaxies well trace the matter distribution in cosmologically large scales. The galaxy density map with mass weight works best as expected from observations that total stellar masses of cluster member galaxies could be a good proxy for cluster masses (e.g. \citealt{per18,pal20}) and that the mass-weighted galaxy number density maps of clusters show good correspondence to the weak-lensing maps (e.g. Fig. 5 of \citealt{hwa2014}). It would be interesting to make a comparison as in this study, but focusing on the detailed features that might have been missed in our method (e.g. Yoo et al., in preparation). Also, this study will be very helpful for understanding the results
  of the mass distribution in and around galaxy clusters
  from wide-field surveys with various tracers, which include 
  DESI (galaxies; \citealt{desi2016}), 
  eROSITA (hot gas; \citealt{predehl2021}), and 
  the Vera C. Rubin Observatory (dark matter and intracluster stars; \citealt{lsst09}).

\section*{Acknowledgements}
We thank the referee for the constructive and detailed comments that helped us to improve the paper. We thank the IllustrisTNG collaboration for making their simulation data publicly available. JHS acknowledges support from the National Research Foundation of Korea grant (2021R1C1C1003785) funded by the Ministry of Science, ICT \& Future Planning.
HSH acknowledges the support by the National Research Foundation of Korea (NRF) grant funded by the Korea government (MSIT) (No. 2021R1A2C1094577). HS was supported by the Basic Science Research Program through the National Research Foundation of Korea (NRF) funded by the Ministry of Education (2020R1I1A1A01069228). JWK acknowledges support from the National Research Foundation of Korea (NRF), grant No. NRF-2019R1C1C1002796, funded by the Korean government (MSIT). 
\bibliography{main.bib}{}

\begin{thebibliography}{}
\expandafter\ifx\csname natexlab\endcsname\relax\def\natexlab#1{#1}\fi
\providecommand{\url}[1]{\href{#1}{#1}}
\providecommand{\dodoi}[1]{doi:~\href{http://doi.org/#1}{\nolinkurl{#1}}}
\providecommand{\doeprint}[1]{\href{http://ascl.net/#1}{\nolinkurl{http://ascl.net/#1}}}
\providecommand{\doarXiv}[1]{\href{https://arxiv.org/abs/#1}{\nolinkurl{https://arxiv.org/abs/#1}}}

\bibitem[{{Allen} {et~al.}(2011){Allen}, {Evrard}, \& {Mantz}}]{all2011}
{Allen}, S.~W., {Evrard}, A.~E., \& {Mantz}, A.~B. 2011, \araa, 49, 409,
  \dodoi{10.1146/annurev-astro-081710-102514}

\bibitem[{{Alonso Asensio} {et~al.}(2020){Alonso Asensio}, {Dalla Vecchia},
  {Bah{\'e}}, {Barnes}, \& {Kay}}]{alo2020}
{Alonso Asensio}, I., {Dalla Vecchia}, C., {Bah{\'e}}, Y.~M., {Barnes}, D.~J.,
  \& {Kay}, S.~T. 2020, \mnras, 494, 1859, \dodoi{10.1093/mnras/staa861}

\bibitem[{{Bershady} {et~al.}(2000){Bershady}, {Jangren}, \&
  {Conselice}}]{ber00}
{Bershady}, M.~A., {Jangren}, A., \& {Conselice}, C.~J. 2000, \aj, 119, 2645,
  \dodoi{10.1086/301386}

\bibitem[{{Bocquet} {et~al.}(2019){Bocquet}, {Dietrich}, {Schrabback}, {Bleem},
  {Klein}, {Allen}, {Applegate}, {Ashby}, {Bautz}, {Bayliss}, {Benson},
  {Brodwin}, {Bulbul}, {Canning}, {Capasso}, {Carlstrom}, {Chang}, {Chiu},
  {Cho}, {Clocchiatti}, {Crawford}, {Crites}, {de Haan}, {Desai}, {Dobbs},
  {Foley}, {Forman}, {Garmire}, {George}, {Gladders}, {Gonzalez}, {Grandis},
  {Gupta}, {Halverson}, {Hlavacek-Larrondo}, {Hoekstra}, {Holder}, {Holzapfel},
  {Hou}, {Hrubes}, {Huang}, {Jones}, {Khullar}, {Knox}, {Kraft}, {Lee}, {von
  der Linden}, {Luong-Van}, {Mantz}, {Marrone}, {McDonald}, {McMahon}, {Meyer},
  {Mocanu}, {Mohr}, {Morris}, {Padin}, {Patil}, {Pryke}, {Rapetti},
  {Reichardt}, {Rest}, {Ruhl}, {Saliwanchik}, {Saro}, {Sayre}, {Schaffer},
  {Shirokoff}, {Stalder}, {Stanford}, {Staniszewski}, {Stark}, {Story},
  {Strazzullo}, {Stubbs}, {Vanderlinde}, {Vieira}, {Vikhlinin}, {Williamson},
  \& {Zenteno}}]{boc2019}
{Bocquet}, S., {Dietrich}, J.~P., {Schrabback}, T., {et~al.} 2019, \apj, 878,
  55, \dodoi{10.3847/1538-4357/ab1f10}

\bibitem[{{B{\"o}hringer}(2002)}]{boh2002}
{B{\"o}hringer}, H. 2002, \ssr, 100, 49, \dodoi{10.1023/A:1015805808907}

\bibitem[{{Bullock} {et~al.}(2001){Bullock}, {Kolatt}, {Sigad}, {Somerville},
  {Kravtsov}, {Klypin}, {Primack}, \& {Dekel}}]{bul2001}
{Bullock}, J.~S., {Kolatt}, T.~S., {Sigad}, Y., {et~al.} 2001, \mnras, 321,
  559, \dodoi{10.1046/j.1365-8711.2001.04068.x}

\bibitem[{{Buote} \& {Tsai}(1995)}]{buo95}
{Buote}, D.~A., \& {Tsai}, J.~C. 1995, \apj, 452, 522, \dodoi{10.1086/176326}

\bibitem[{{Chandrasekhar}(1961)}]{cha91}
{Chandrasekhar}, S. 1961, {Hydrodynamic and hydromagnetic stability}

\bibitem[{{Clowe} {et~al.}(2004){Clowe}, {Gonzalez}, \& {Markevitch}}]{clo04}
{Clowe}, D., {Gonzalez}, A., \& {Markevitch}, M. 2004, \apj, 604, 596,
  \dodoi{10.1086/381970}

\bibitem[{{Clowe} {et~al.}(2012){Clowe}, {Markevitch}, {Brada{\v{c}}},
  {Gonzalez}, {Chung}, {Massey}, \& {Zaritsky}}]{clo2012}
{Clowe}, D., {Markevitch}, M., {Brada{\v{c}}}, M., {et~al.} 2012, \apj, 758,
  128, \dodoi{10.1088/0004-637X/758/2/128}

\bibitem[{{Cui} {et~al.}(2017){Cui}, {Power}, {Borgani}, {Knebe}, {Lewis},
  {Murante}, \& {Poole}}]{cui17}
{Cui}, W., {Power}, C., {Borgani}, S., {et~al.} 2017, \mnras, 464, 2502,
  \dodoi{10.1093/mnras/stw2567}

\bibitem[{{DESI Collaboration} {et~al.}(2016){DESI Collaboration}, {Aghamousa},
  {Aguilar}, {Ahlen}, {Alam}, {Allen}, {Allende Prieto}, {Annis}, {Bailey},
  {Balland}, {Ballester}, {Baltay}, {Beaufore}, {Bebek}, {Beers}, {Bell},
  {Bernal}, {Besuner}, {Beutler}, {Blake}, {Bleuler}, {Blomqvist}, {Blum},
  {Bolton}, {Briceno}, {Brooks}, {Brownstein}, {Buckley-Geer}, {Burden},
  {Burtin}, {Busca}, {Cahn}, {Cai}, {Cardiel-Sas}, {Carlberg}, {Carton},
  {Casas}, {Castander}, {Cervantes-Cota}, {Claybaugh}, {Close}, {Coker},
  {Cole}, {Comparat}, {Cooper}, {Cousinou}, {Crocce}, {Cuby}, {Cunningham},
  {Davis}, {Dawson}, {de la Macorra}, {De Vicente}, {Delubac}, {Derwent},
  {Dey}, {Dhungana}, {Ding}, {Doel}, {Duan}, {Ealet}, {Edelstein},
  {Eftekharzadeh}, {Eisenstein}, {Elliott}, {Escoffier}, {Evatt}, {Fagrelius},
  {Fan}, {Fanning}, {Farahi}, {Farihi}, {Favole}, {Feng}, {Fernandez},
  {Findlay}, {Finkbeiner}, {Fitzpatrick}, {Flaugher}, {Flender}, {Font-Ribera},
  {Forero-Romero}, {Fosalba}, {Frenk}, {Fumagalli}, {Gaensicke}, {Gallo},
  {Garcia-Bellido}, {Gaztanaga}, {Pietro Gentile Fusillo}, {Gerard},
  {Gershkovich}, {Giannantonio}, {Gillet}, {Gonzalez-de-Rivera},
  {Gonzalez-Perez}, {Gott}, {Graur}, {Gutierrez}, {Guy}, {Habib}, {Heetderks},
  {Heetderks}, {Heitmann}, {Hellwing}, {Herrera}, {Ho}, {Holland}, {Honscheid},
  {Huff}, {Hutchinson}, {Huterer}, {Hwang}, {Illa Laguna}, {Ishikawa},
  {Jacobs}, {Jeffrey}, {Jelinsky}, {Jennings}, {Jiang}, {Jimenez}, {Johnson},
  {Joyce}, {Jullo}, {Juneau}, {Kama}, {Karcher}, {Karkar}, {Kehoe}, {Kennamer},
  {Kent}, {Kilbinger}, {Kim}, {Kirkby}, {Kisner}, {Kitanidis}, {Kneib},
  {Koposov}, {Kovacs}, {Koyama}, {Kremin}, {Kron}, {Kronig}, {Kueter-Young},
  {Lacey}, {Lafever}, {Lahav}, {Lambert}, {Lampton}, {Landriau}, {Lang},
  {Lauer}, {Le Goff}, {Le Guillou}, {Le Van Suu}, {Lee}, {Lee}, {Leitner},
  {Lesser}, {Levi}, {L'Huillier}, {Li}, {Liang}, {Lin}, {Linder}, {Loebman},
  {Luki{\'c}}, {Ma}, {MacCrann}, {Magneville}, {Makarem}, {Manera}, {Manser},
  {Marshall}, {Martini}, {Massey}, {Matheson}, {McCauley}, {McDonald},
  {McGreer}, {Meisner}, {Metcalfe}, {Miller}, {Miquel}, {Moustakas}, {Myers},
  {Naik}, {Newman}, {Nichol}, {Nicola}, {Nicolati da Costa}, {Nie}, {Niz},
  {Norberg}, {Nord}, {Norman}, {Nugent}, {O'Brien}, {Oh}, {Olsen}, {Padilla},
  {Padmanabhan}, {Padmanabhan}, {Palanque-Delabrouille}, {Palmese},
  {Pappalardo}, {P{\^a}ris}, {Park}, {Patej}, {Peacock}, {Peiris}, {Peng},
  {Percival}, {Perruchot}, {Pieri}, {Pogge}, {Pollack}, {Poppett}, {Prada},
  {Prakash}, {Probst}, {Rabinowitz}, {Raichoor}, {Ree}, {Refregier}, {Regal},
  {Reid}, {Reil}, {Rezaie}, {Rockosi}, {Roe}, {Ronayette}, {Roodman}, {Ross},
  {Ross}, {Rossi}, {Rozo}, {Ruhlmann-Kleider}, {Rykoff}, {Sabiu}, {Samushia},
  {Sanchez}, {Sanchez}, {Schlegel}, {Schneider}, {Schubnell}, {Secroun},
  {Seljak}, {Seo}, {Serrano}, {Shafieloo}, {Shan}, {Sharples}, {Sholl},
  {Shourt}, {Silber}, {Silva}, {Sirk}, {Slosar}, {Smith}, {Smoot}, {Som},
  {Song}, {Sprayberry}, {Staten}, {Stefanik}, {Tarle}, {Sien Tie}, {Tinker},
  {Tojeiro}, {Valdes}, {Valenzuela}, {Valluri}, {Vargas-Magana}, {Verde},
  {Walker}, {Wang}, {Wang}, {Weaver}, {Weaverdyck}, {Wechsler}, {Weinberg},
  {White}, {Yang}, {Yeche}, {Zhang}, {Zhao}, {Zheng}, {Zhou}, {Zhou}, {Zhu},
  {Zou}, \& {Zu}}]{desi2016}
{DESI Collaboration}, {Aghamousa}, A., {Aguilar}, J., {et~al.} 2016, arXiv
  e-prints, arXiv:1611.00036.
\newblock \doarXiv{1611.00036}

\bibitem[{{Desjacques} {et~al.}(2018){Desjacques}, {Jeong}, \&
  {Schmidt}}]{des1028}
{Desjacques}, V., {Jeong}, D., \& {Schmidt}, F. 2018, \physrep, 733, 1,
  \dodoi{10.1016/j.physrep.2017.12.002}

\bibitem[{{Dodelson}(2004)}]{dod04}
{Dodelson}, S. 2004, \prd, 70, 023008, \dodoi{10.1103/PhysRevD.70.023008}

\bibitem[{{Ettori} {et~al.}(2013){Ettori}, {Donnarumma}, {Pointecouteau},
  {Reiprich}, {Giodini}, {Lovisari}, \& {Schmidt}}]{ett2013}
{Ettori}, S., {Donnarumma}, A., {Pointecouteau}, E., {et~al.} 2013, \ssr, 177,
  119, \dodoi{10.1007/s11214-013-9976-7}

\bibitem[{{Feldmeier} {et~al.}(2004){Feldmeier}, {Ciardullo}, {Jacoby}, \&
  {Durrell}}]{fel2004}
{Feldmeier}, J.~J., {Ciardullo}, R., {Jacoby}, G.~H., \& {Durrell}, P.~R. 2004,
  \apj, 615, 196, \dodoi{10.1086/424372}

\bibitem[{{Geller} {et~al.}(2013){Geller}, {Diaferio}, {Rines}, \&
  {Serra}}]{gel2013}
{Geller}, M.~J., {Diaferio}, A., {Rines}, K.~J., \& {Serra}, A.~L. 2013, \apj,
  764, 58, \dodoi{10.1088/0004-637X/764/1/58}

\bibitem[{{Geller} {et~al.}(2014){Geller}, {Hwang}, {Diaferio}, {Kurtz}, {Coe},
  \& {Rines}}]{gel2014}
{Geller}, M.~J., {Hwang}, H.~S., {Diaferio}, A., {et~al.} 2014, \apj, 783, 52,
  \dodoi{10.1088/0004-637X/783/1/52}

\bibitem[{{Hernquist}(1990)}]{her90}
{Hernquist}, L. 1990, \apj, 356, 359, \dodoi{10.1086/168845}

\bibitem[{{Hoekstra}(2001)}]{hoe01}
{Hoekstra}, H. 2001, \aap, 370, 743, \dodoi{10.1051/0004-6361:20010293}

\bibitem[{{Hoffer} {et~al.}(2012){Hoffer}, {Donahue}, {Hicks}, \&
  {Barthelemy}}]{hof12}
{Hoffer}, A.~S., {Donahue}, M., {Hicks}, A., \& {Barthelemy}, R.~S. 2012,
  \apjs, 199, 23, \dodoi{10.1088/0067-0049/199/1/23}

\bibitem[{{Holland} {et~al.}(2015){Holland}, {B{\"o}hringer}, {Chon}, \&
  {Pierini}}]{hol15}
{Holland}, J.~G., {B{\"o}hringer}, H., {Chon}, G., \& {Pierini}, D. 2015,
  \mnras, 448, 2644, \dodoi{10.1093/mnras/stv097}

\bibitem[{{Hong} {et~al.}(2021){Hong}, {Jeong}, {Hwang}, \& {Kim}}]{hon21}
{Hong}, S.~E., {Jeong}, D., {Hwang}, H.~S., \& {Kim}, J. 2021, \apj, 913, 76,
  \dodoi{10.3847/1538-4357/abf040}

\bibitem[{{Hudson} {et~al.}(2010){Hudson}, {Mittal}, {Reiprich}, {Nulsen},
  {Andernach}, \& {Sarazin}}]{hud10}
{Hudson}, D.~S., {Mittal}, R., {Reiprich}, T.~H., {et~al.} 2010, \aap, 513,
  A37, \dodoi{10.1051/0004-6361/200912377}

\bibitem[{{Hwang} {et~al.}(2014){Hwang}, {Geller}, {Diaferio}, {Rines}, \&
  {Zahid}}]{hwa2014}
{Hwang}, H.~S., {Geller}, M.~J., {Diaferio}, A., {Rines}, K.~J., \& {Zahid},
  H.~J. 2014, \apj, 797, 106, \dodoi{10.1088/0004-637X/797/2/106}

\bibitem[{{Jee} {et~al.}(2014){Jee}, {Hoekstra}, {Mahdavi}, \&
  {Babul}}]{jee2014}
{Jee}, M.~J., {Hoekstra}, H., {Mahdavi}, A., \& {Babul}, A. 2014, \apj, 783,
  78, \dodoi{10.1088/0004-637X/783/2/78}

\bibitem[{{Kaiser}(1984)}]{kai1984}
{Kaiser}, N. 1984, \apjl, 284, L9, \dodoi{10.1086/184341}

\bibitem[{{Katayama} {et~al.}(2003){Katayama}, {Hayashida}, {Takahara}, \&
  {Fujita}}]{kat03}
{Katayama}, H., {Hayashida}, K., {Takahara}, F., \& {Fujita}, Y. 2003, \apj,
  585, 687, \dodoi{10.1086/346126}

\bibitem[{{Kim} {et~al.}(2022){Kim}, {Ko}, {Smith}, {Kim}, {Hwang}, {Song},
  {Shin}, \& {Yoo}}]{kim22}
{Kim}, H., {Ko}, J., {Smith}, R., {et~al.} 2022, \apj, 928, 170,
  \dodoi{10.3847/1538-4357/ac510e}

\bibitem[{{Kim} {et~al.}(2017){Kim}, {Ko}, {Hwang}, {Edge}, {Lee}, {Lee}, \&
  {Jeong}}]{kim17}
{Kim}, J.-W., {Ko}, J., {Hwang}, H.~S., {et~al.} 2017, \apj, 836, 105,
  \dodoi{10.3847/1538-4357/aa5b8e}

\bibitem[{{Ko} \& {Jee}(2018)}]{ko2018}
{Ko}, J., \& {Jee}, M.~J. 2018, \apj, 862, 95, \dodoi{10.3847/1538-4357/aacbda}

\bibitem[{{Ko} {et~al.}(2017){Ko}, {Hwang}, {Lee}, {Park}, {Lim}, {Sohn},
  {Jang}, {Hwang}, \& {Park}}]{ko2017}
{Ko}, Y., {Hwang}, H.~S., {Lee}, M.~G., {et~al.} 2017, \apj, 835, 212,
  \dodoi{10.3847/1538-4357/835/2/212}

\bibitem[{{Lee} {et~al.}(2010){Lee}, {Park}, \& {Hwang}}]{lee2010}
{Lee}, M.~G., {Park}, H.~S., \& {Hwang}, H.~S. 2010, Science, 328, 334,
  \dodoi{10.1126/science.1186496}

\bibitem[{{Liu} {et~al.}(2018){Liu}, {Yu}, {Diaferio}, {Tozzi}, {Hwang},
  {Umetsu}, {Okabe}, \& {Yang}}]{liu2018}
{Liu}, A., {Yu}, H., {Diaferio}, A., {et~al.} 2018, \apj, 863, 102,
  \dodoi{10.3847/1538-4357/aad090}

\bibitem[{{Longobardi} {et~al.}(2018){Longobardi}, {Arnaboldi}, {Gerhard},
  {Pulsoni}, \& {S{\"o}ldner-Rembold}}]{lon2018}
{Longobardi}, A., {Arnaboldi}, M., {Gerhard}, O., {Pulsoni}, C., \&
  {S{\"o}ldner-Rembold}, I. 2018, \aap, 620, A111,
  \dodoi{10.1051/0004-6361/201832729}

\bibitem[{{Lotz} {et~al.}(2004){Lotz}, {Primack}, \& {Madau}}]{lot04}
{Lotz}, J.~M., {Primack}, J., \& {Madau}, P. 2004, \aj, 128, 163,
  \dodoi{10.1086/421849}

\bibitem[{{LSST Science Collaboration} {et~al.}(2009){LSST Science
  Collaboration}, {Abell}, {Allison}, {Anderson}, {Andrew}, {Angel}, {Armus},
  {Arnett}, {Asztalos}, {Axelrod}, {Bailey}, {Ballantyne}, {Bankert},
  {Barkhouse}, {Barr}, {Barrientos}, {Barth}, {Bartlett}, {Becker}, {Becla},
  {Beers}, {Bernstein}, {Biswas}, {Blanton}, {Bloom}, {Bochanski}, {Boeshaar},
  {Borne}, {Bradac}, {Brandt}, {Bridge}, {Brown}, {Brunner}, {Bullock},
  {Burgasser}, {Burge}, {Burke}, {Cargile}, {Chandrasekharan}, {Chartas},
  {Chesley}, {Chu}, {Cinabro}, {Claire}, {Claver}, {Clowe}, {Connolly}, {Cook},
  {Cooke}, {Cooray}, {Covey}, {Culliton}, {de Jong}, {de Vries}, {Debattista},
  {Delgado}, {Dell'Antonio}, {Dhital}, {Di Stefano}, {Dickinson}, {Dilday},
  {Djorgovski}, {Dobler}, {Donalek}, {Dubois-Felsmann}, {Durech},
  {Eliasdottir}, {Eracleous}, {Eyer}, {Falco}, {Fan}, {Fassnacht}, {Ferguson},
  {Fernandez}, {Fields}, {Finkbeiner}, {Figueroa}, {Fox}, {Francke}, {Frank},
  {Frieman}, {Fromenteau}, {Furqan}, {Galaz}, {Gal-Yam}, {Garnavich},
  {Gawiser}, {Geary}, {Gee}, {Gibson}, {Gilmore}, {Grace}, {Green}, {Gressler},
  {Grillmair}, {Habib}, {Haggerty}, {Hamuy}, {Harris}, {Hawley}, {Heavens},
  {Hebb}, {Henry}, {Hileman}, {Hilton}, {Hoadley}, {Holberg}, {Holman},
  {Howell}, {Infante}, {Ivezic}, {Jacoby}, {Jain}, {R}, {Jedicke}, {Jee},
  {Garrett Jernigan}, {Jha}, {Johnston}, {Jones}, {Juric}, {Kaasalainen},
  {Styliani}, {Kafka}, {Kahn}, {Kaib}, {Kalirai}, {Kantor}, {Kasliwal},
  {Keeton}, {Kessler}, {Knezevic}, {Kowalski}, {Krabbendam}, {Krughoff},
  {Kulkarni}, {Kuhlman}, {Lacy}, {Lepine}, {Liang}, {Lien}, {Lira}, {Long},
  {Lorenz}, {Lotz}, {Lupton}, {Lutz}, {Macri}, {Mahabal}, {Mandelbaum},
  {Marshall}, {May}, {McGehee}, {Meadows}, {Meert}, {Milani}, {Miller},
  {Miller}, {Mills}, {Minniti}, {Monet}, {Mukadam}, {Nakar}, {Neill}, {Newman},
  {Nikolaev}, {Nordby}, {O'Connor}, {Oguri}, {Oliver}, {Olivier}, {Olsen},
  {Olsen}, {Olszewski}, {Oluseyi}, {Padilla}, {Parker}, {Pepper}, {Peterson},
  {Petry}, {Pinto}, {Pizagno}, {Popescu}, {Prsa}, {Radcka}, {Raddick},
  {Rasmussen}, {Rau}, {Rho}, {Rhoads}, {Richards}, {Ridgway}, {Robertson},
  {Roskar}, {Saha}, {Sarajedini}, {Scannapieco}, {Schalk}, {Schindler},
  {Schmidt}, {Schmidt}, {Schneider}, {Schumacher}, {Scranton}, {Sebag},
  {Seppala}, {Shemmer}, {Simon}, {Sivertz}, {Smith}, {Allyn Smith}, {Smith},
  {Spitz}, {Stanford}, {Stassun}, {Strader}, {Strauss}, {Stubbs}, {Sweeney},
  {Szalay}, {Szkody}, {Takada}, {Thorman}, {Trilling}, {Trimble}, {Tyson}, {Van
  Berg}, {Vanden Berk}, {VanderPlas}, {Verde}, {Vrsnak}, {Walkowicz},
  {Wandelt}, {Wang}, {Wang}, {Warner}, {Wechsler}, {West}, {Wiecha},
  {Williams}, {Willman}, {Wittman}, {Wolff}, {Wood-Vasey}, {Wozniak}, {Young},
  {Zentner}, \& {Zhan}}]{lsst09}
{LSST Science Collaboration}, {Abell}, P.~A., {Allison}, J., {et~al.} 2009,
  arXiv e-prints, arXiv:0912.0201.
\newblock \doarXiv{0912.0201}

\bibitem[{{Markwardt}(2009)}]{mar09}
{Markwardt}, C.~B. 2009, in Astronomical Society of the Pacific Conference
  Series, Vol. 411, Astronomical Data Analysis Software and Systems XVIII, ed.
  D.~A. {Bohlender}, D.~{Durand}, \& P.~{Dowler}, 251.
\newblock \doarXiv{0902.2850}

\bibitem[{{Massey} {et~al.}(2015){Massey}, {Williams}, {Smit}, {Swinbank},
  {Kitching}, {Harvey}, {Jauzac}, {Israel}, {Clowe}, {Edge}, {Hilton}, {Jullo},
  {Leonard}, {Liesenborgs}, {Merten}, {Mohammed}, {Nagai}, {Richard},
  {Robertson}, {Saha}, {Santana}, {Stott}, \& {Tittley}}]{mas15}
{Massey}, R., {Williams}, L., {Smit}, R., {et~al.} 2015, \mnras, 449, 3393,
  \dodoi{10.1093/mnras/stv467}

\bibitem[{{Mihos} {et~al.}(2005){Mihos}, {Harding}, {Feldmeier}, \&
  {Morrison}}]{mih2005}
{Mihos}, J.~C., {Harding}, P., {Feldmeier}, J., \& {Morrison}, H. 2005, \apjl,
  631, L41, \dodoi{10.1086/497030}

\bibitem[{{Mohr} {et~al.}(1993){Mohr}, {Fabricant}, \& {Geller}}]{moh93}
{Mohr}, J.~J., {Fabricant}, D.~G., \& {Geller}, M.~J. 1993, \apj, 413, 492,
  \dodoi{10.1086/173019}

\bibitem[{{Montes} \& {Trujillo}(2019)}]{mon2019}
{Montes}, M., \& {Trujillo}, I. 2019, \mnras, 482, 2838,
  \dodoi{10.1093/mnras/sty2858}

\bibitem[{{Moura} {et~al.}(2021){Moura}, {Machado}, \&
  {Monteiro-Oliveira}}]{mou21}
{Moura}, M.~T., {Machado}, R. E.~G., \& {Monteiro-Oliveira}, R. 2021, \mnras,
  500, 1858, \dodoi{10.1093/mnras/staa3399}

\bibitem[{{Navarro} {et~al.}(1996){Navarro}, {Frenk}, \& {White}}]{nfw96}
{Navarro}, J.~F., {Frenk}, C.~S., \& {White}, S. D.~M. 1996, \apj, 462, 563,
  \dodoi{10.1086/177173}

\bibitem[{{Nelson} {et~al.}(2019){Nelson}, {Springel}, {Pillepich},
  {Rodriguez-Gomez}, {Torrey}, {Genel}, {Vogelsberger}, {Pakmor}, {Marinacci},
  {Weinberger}, {Kelley}, {Lovell}, {Diemer}, \& {Hernquist}}]{nel19}
{Nelson}, D., {Springel}, V., {Pillepich}, A., {et~al.} 2019, Computational
  Astrophysics and Cosmology, 6, 2, \dodoi{10.1186/s40668-019-0028-x}

\bibitem[{{Okabe} {et~al.}(2014){Okabe}, {Futamase}, {Kajisawa}, \&
  {Kuroshima}}]{oka2014}
{Okabe}, N., {Futamase}, T., {Kajisawa}, M., \& {Kuroshima}, R. 2014, \apj,
  784, 90, \dodoi{10.1088/0004-637X/784/2/90}

\bibitem[{{Okabe} {et~al.}(2010){Okabe}, {Takada}, {Umetsu}, {Futamase}, \&
  {Smith}}]{oka2010}
{Okabe}, N., {Takada}, M., {Umetsu}, K., {Futamase}, T., \& {Smith}, G.~P.
  2010, \pasj, 62, 811, \dodoi{10.1093/pasj/62.3.811}

\bibitem[{{Palmese} {et~al.}(2020){Palmese}, {Annis}, {Burgad}, {Farahi},
  {Soares-Santos}, {Welch}, {da Silva Pereira}, {Lin}, {Bhargava}, {Hollowood},
  {Wilkinson}, {Giles}, {Jeltema}, {Romer}, {Evrard}, {Hilton}, {Vergara
  Cervantes}, {Bermeo}, {Mayers}, {DeRose}, {Gruen}, {Hartley}, {Lahav},
  {Leistedt}, {McClintock}, {Rozo}, {Rykoff}, {Varga}, {Wechsler}, {Zhang},
  {Avila}, {Brooks}, {Buckley-Geer}, {Burke}, {Carnero Rosell}, {Carrasco
  Kind}, {Carretero}, {Castander}, {Collins}, {da Costa}, {Desai}, {De
  Vicente}, {Diehl}, {Dietrich}, {Doel}, {Flaugher}, {Fosalba}, {Frieman},
  {Garc{\'\i}a-Bellido}, {Gerdes}, {Gruendl}, {Gschwend}, {Gutierrez},
  {Honscheid}, {James}, {Krause}, {Kuehn}, {Kuropatkin}, {Liddle}, {Lima},
  {Maia}, {Mann}, {Marshall}, {Menanteau}, {Miquel}, {Ogando}, {Plazas},
  {Roodman}, {Rooney}, {Sahlen}, {Sanchez}, {Scarpine}, {Schubnell}, {Serrano},
  {Sevilla-Noarbe}, {Sobreira}, {Stott}, {Suchyta}, {Swanson}, {Tarle},
  {Thomas}, {Tucker}, {Viana}, {Vikram}, {Walker}, \& {DES
  Collaboration}}]{pal20}
{Palmese}, A., {Annis}, J., {Burgad}, J., {et~al.} 2020, \mnras, 493, 4591,
  \dodoi{10.1093/mnras/staa526}

\bibitem[{{Parekh} {et~al.}(2015){Parekh}, {van der Heyden}, {Ferrari},
  {Angus}, \& {Holwerda}}]{par15}
{Parekh}, V., {van der Heyden}, K., {Ferrari}, C., {Angus}, G., \& {Holwerda},
  B. 2015, \aap, 575, A127, \dodoi{10.1051/0004-6361/201424123}

\bibitem[{{Pereira} {et~al.}(2018){Pereira}, {Soares-Santos}, {Makler},
  {Annis}, {Lin}, {Palmese}, {Vitorelli}, {Welch}, {Caminha}, {Erben},
  {Moraes}, \& {Shan}}]{per18}
{Pereira}, M. E.~S., {Soares-Santos}, M., {Makler}, M., {et~al.} 2018, \mnras,
  474, 1361, \dodoi{10.1093/mnras/stx2831}

\bibitem[{{Pillepich} {et~al.}(2018{\natexlab{a}}){Pillepich}, {Nelson},
  {Hernquist}, {Springel}, {Pakmor}, {Torrey}, {Weinberger}, {Genel}, {Naiman},
  {Marinacci}, \& {Vogelsberger}}]{pil18b}
{Pillepich}, A., {Nelson}, D., {Hernquist}, L., {et~al.} 2018{\natexlab{a}},
  \mnras, 475, 648, \dodoi{10.1093/mnras/stx3112}

\bibitem[{{Pillepich} {et~al.}(2018{\natexlab{b}}){Pillepich}, {Springel},
  {Nelson}, {Genel}, {Naiman}, {Pakmor}, {Hernquist}, {Torrey}, {Vogelsberger},
  {Weinberger}, \& {Marinacci}}]{pil18a}
{Pillepich}, A., {Springel}, V., {Nelson}, D., {et~al.} 2018{\natexlab{b}},
  \mnras, 473, 4077, \dodoi{10.1093/mnras/stx2656}

\bibitem[{{Planck Collaboration} {et~al.}(2013){Planck Collaboration}, {Ade},
  {Aghanim}, {Arnaud}, {Ashdown}, {Atrio-Barandela}, {Aumont}, {Baccigalupi},
  {Balbi}, {Banday}, {Barreiro}, {Bartlett}, {Battaner}, {Benabed},
  {Beno{\^\i}t}, {Bernard}, {Bersanelli}, {Bhatia}, {Bikmaev}, {Bobin},
  {B{\"o}hringer}, {Bonaldi}, {Bond}, {Borgani}, {Borrill}, {Bouchet},
  {Bourdin}, {Brown}, {Burenin}, {Burigana}, {Cabella}, {Cardoso}, {Carvalho},
  {Castex}, {Catalano}, {Cay{\'o}n}, {Chamballu}, {Chiang}, {Chon},
  {Christensen}, {Churazov}, {Clements}, {Colafrancesco}, {Colombi}, {Colombo},
  {Comis}, {Coulais}, {Crill}, {Cuttaia}, {Da Silva}, {Dahle}, {Danese},
  {Davis}, {de Bernardis}, {de Gasperis}, {de Zotti}, {Delabrouille},
  {D{\'e}mocl{\`e}s}, {D{\'e}sert}, {Diego}, {Dolag}, {Dole}, {Donzelli},
  {Dor{\'e}}, {D{\"o}rl}, {Douspis}, {Dupac}, {Efstathiou}, {En{\ss}lin},
  {Eriksen}, {Finelli}, {Flores-Cacho}, {Forni}, {Fosalba}, {Frailis},
  {Franceschi}, {Frommert}, {Galeotta}, {Ganga}, {G{\'e}nova-Santos}, {Giard},
  {Giraud-H{\'e}raud}, {Gonz{\'a}lez-Nuevo}, {G{\'o}rski}, {Gregorio},
  {Gruppuso}, {Hansen}, {Harrison}, {Hempel}, {Henrot-Versill{\'e}},
  {Hern{\'a}ndez-Monteagudo}, {Herranz}, {Hildebrandt}, {Hivon}, {Hobson},
  {Holmes}, {Hurier}, {Jaffe}, {Jaffe}, {Jagemann}, {Jones}, {Juvela},
  {Keih{\"a}nen}, {Khamitov}, {Kisner}, {Kneissl}, {Knoche}, {Knox}, {Kunz},
  {Kurki-Suonio}, {Lagache}, {L{\"a}hteenm{\"a}ki}, {Lamarre}, {Lasenby},
  {Lawrence}, {Le Jeune}, {Leonardi}, {Liddle}, {Lilje}, {L{\'o}pez-Caniego},
  {Luzzi}, {Mac{\'\i}as-P{\'e}rez}, {Maino}, {Mandolesi}, {Maris}, {Marleau},
  {Marshall}, {Mart{\'\i}nez-Gonz{\'a}lez}, {Masi}, {Massardi}, {Matarrese},
  {Mazzotta}, {Mei}, {Melchiorri}, {Melin}, {Mendes}, {Mennella}, {Mitra},
  {Miville-Desch{\^e}nes}, {Moneti}, {Montier}, {Morgante}, {Mortlock},
  {Munshi}, {Murphy}, {Naselsky}, {Nati}, {Natoli}, {N{\o}rgaard-Nielsen},
  {Noviello}, {Novikov}, {Novikov}, {Osborne}, {Pajot}, {Paoletti}, {Pasian},
  {Patanchon}, {Perdereau}, {Perotto}, {Perrotta}, {Piacentini}, {Piat},
  {Pierpaoli}, {Piffaretti}, {Plaszczynski}, {Pointecouteau}, {Polenta},
  {Ponthieu}, {Popa}, {Poutanen}, {Pratt}, {Prunet}, {Puget}, {Rachen},
  {Reach}, {Rebolo}, {Reinecke}, {Remazeilles}, {Renault}, {Ricciardi},
  {Riller}, {Ristorcelli}, {Rocha}, {Roman}, {Rosset}, {Rossetti},
  {Rubi{\~n}o-Mart{\'\i}n}, {Rusholme}, {Sandri}, {Savini}, {Scott}, {Smoot},
  {Starck}, {Sudiwala}, {Sunyaev}, {Sutton}, {Suur-Uski}, {Sygnet}, {Tauber},
  {Terenzi}, {Toffolatti}, {Tomasi}, {Tristram}, {Tuovinen}, {Valenziano}, {Van
  Tent}, {Varis}, {Vielva}, {Villa}, {Vittorio}, {Wade}, {Wandelt}, {Welikala},
  {White}, {White}, {Yvon}, {Zacchei}, \& {Zonca}}]{Planck13}
{Planck Collaboration}, {Ade}, P.~A.~R., {Aghanim}, N., {et~al.} 2013, \aap,
  550, A131, \dodoi{10.1051/0004-6361/201220040}

\bibitem[{{Planck Collaboration} {et~al.}(2016){Planck Collaboration}, {Ade},
  {Aghanim}, {Arnaud}, {Ashdown}, {Aumont}, {Baccigalupi}, {Banday},
  {Barreiro}, {Bartlett}, {Bartolo}, {Battaner}, {Battye}, {Benabed},
  {Beno{\^\i}t}, {Benoit-L{\'e}vy}, {Bernard}, {Bersanelli}, {Bielewicz},
  {Bock}, {Bonaldi}, {Bonavera}, {Bond}, {Borrill}, {Bouchet}, {Boulanger},
  {Bucher}, {Burigana}, {Butler}, {Calabrese}, {Cardoso}, {Catalano},
  {Challinor}, {Chamballu}, {Chary}, {Chiang}, {Chluba}, {Christensen},
  {Church}, {Clements}, {Colombi}, {Colombo}, {Combet}, {Coulais}, {Crill},
  {Curto}, {Cuttaia}, {Danese}, {Davies}, {Davis}, {de Bernardis}, {de Rosa},
  {de Zotti}, {Delabrouille}, {D{\'e}sert}, {Di Valentino}, {Dickinson},
  {Diego}, {Dolag}, {Dole}, {Donzelli}, {Dor{\'e}}, {Douspis}, {Ducout},
  {Dunkley}, {Dupac}, {Efstathiou}, {Elsner}, {En{\ss}lin}, {Eriksen},
  {Farhang}, {Fergusson}, {Finelli}, {Forni}, {Frailis}, {Fraisse},
  {Franceschi}, {Frejsel}, {Galeotta}, {Galli}, {Ganga}, {Gauthier}, {Gerbino},
  {Ghosh}, {Giard}, {Giraud-H{\'e}raud}, {Giusarma}, {Gjerl{\o}w},
  {Gonz{\'a}lez-Nuevo}, {G{\'o}rski}, {Gratton}, {Gregorio}, {Gruppuso},
  {Gudmundsson}, {Hamann}, {Hansen}, {Hanson}, {Harrison}, {Helou},
  {Henrot-Versill{\'e}}, {Hern{\'a}ndez-Monteagudo}, {Herranz}, {Hildebrand t},
  {Hivon}, {Hobson}, {Holmes}, {Hornstrup}, {Hovest}, {Huang}, {Huffenberger},
  {Hurier}, {Jaffe}, {Jaffe}, {Jones}, {Juvela}, {Keih{\"a}nen}, {Keskitalo},
  {Kisner}, {Kneissl}, {Knoche}, {Knox}, {Kunz}, {Kurki-Suonio}, {Lagache},
  {L{\"a}hteenm{\"a}ki}, {Lamarre}, {Lasenby}, {Lattanzi}, {Lawrence}, {Leahy},
  {Leonardi}, {Lesgourgues}, {Levrier}, {Lewis}, {Liguori}, {Lilje},
  {Linden-V{\o}rnle}, {L{\'o}pez-Caniego}, {Lubin}, {Mac{\'\i}as-P{\'e}rez},
  {Maggio}, {Maino}, {Mandolesi}, {Mangilli}, {Marchini}, {Maris}, {Martin},
  {Martinelli}, {Mart{\'\i}nez-Gonz{\'a}lez}, {Masi}, {Matarrese}, {McGehee},
  {Meinhold}, {Melchiorri}, {Melin}, {Mendes}, {Mennella}, {Migliaccio},
  {Millea}, {Mitra}, {Miville-Desch{\^e}nes}, {Moneti}, {Montier}, {Morgante},
  {Mortlock}, {Moss}, {Munshi}, {Murphy}, {Naselsky}, {Nati}, {Natoli},
  {Netterfield}, {N{\o}rgaard-Nielsen}, {Noviello}, {Novikov}, {Novikov},
  {Oxborrow}, {Paci}, {Pagano}, {Pajot}, {Paladini}, {Paoletti}, {Partridge},
  {Pasian}, {Patanchon}, {Pearson}, {Perdereau}, {Perotto}, {Perrotta},
  {Pettorino}, {Piacentini}, {Piat}, {Pierpaoli}, {Pietrobon}, {Plaszczynski},
  {Pointecouteau}, {Polenta}, {Popa}, {Pratt}, {Pr{\'e}zeau}, {Prunet},
  {Puget}, {Rachen}, {Reach}, {Rebolo}, {Reinecke}, {Remazeilles}, {Renault},
  {Renzi}, {Ristorcelli}, {Rocha}, {Rosset}, {Rossetti}, {Roudier},
  {Rouill{\'e} d'Orfeuil}, {Rowan-Robinson}, {Rubi{\~n}o-Mart{\'\i}n},
  {Rusholme}, {Said}, {Salvatelli}, {Salvati}, {Sandri}, {Santos},
  {Savelainen}, {Savini}, {Scott}, {Seiffert}, {Serra}, {Shellard}, {Spencer},
  {Spinelli}, {Stolyarov}, {Stompor}, {Sudiwala}, {Sunyaev}, {Sutton},
  {Suur-Uski}, {Sygnet}, {Tauber}, {Terenzi}, {Toffolatti}, {Tomasi},
  {Tristram}, {Trombetti}, {Tucci}, {Tuovinen}, {T{\"u}rler}, {Umana},
  {Valenziano}, {Valiviita}, {Van Tent}, {Vielva}, {Villa}, {Wade}, {Wandelt},
  {Wehus}, {White}, {White}, {Wilkinson}, {Yvon}, {Zacchei}, \&
  {Zonca}}]{planck16}
---. 2016, \aap, 594, A13, \dodoi{10.1051/0004-6361/201525830}

\bibitem[{{Predehl} {et~al.}(2021){Predehl}, {Andritschke}, {Arefiev},
  {Babyshkin}, {Batanov}, {Becker}, {B{\"o}hringer}, {Bogomolov}, {Boller},
  {Borm}, {Bornemann}, {Br{\"a}uninger}, {Br{\"u}ggen}, {Brunner}, {Brusa},
  {Bulbul}, {Buntov}, {Burwitz}, {Burkert}, {Clerc}, {Churazov}, {Coutinho},
  {Dauser}, {Dennerl}, {Doroshenko}, {Eder}, {Emberger}, {Eraerds},
  {Finoguenov}, {Freyberg}, {Friedrich}, {Friedrich}, {F{\"u}rmetz},
  {Georgakakis}, {Gilfanov}, {Granato}, {Grossberger}, {Gueguen}, {Gureev},
  {Haberl}, {H{\"a}lker}, {Hartner}, {Hasinger}, {Huber}, {Ji}, {Kienlin},
  {Kink}, {Korotkov}, {Kreykenbohm}, {Lamer}, {Lomakin}, {Lapshov}, {Liu},
  {Maitra}, {Meidinger}, {Menz}, {Merloni}, {Mernik}, {Mican}, {Mohr},
  {M{\"u}ller}, {Nandra}, {Nazarov}, {Pacaud}, {Pavlinsky}, {Perinati},
  {Pfeffermann}, {Pietschner}, {Ramos-Ceja}, {Rau}, {Reiffers}, {Reiprich},
  {Robrade}, {Salvato}, {Sanders}, {Santangelo}, {Sasaki}, {Scheuerle},
  {Schmid}, {Schmitt}, {Schwope}, {Shirshakov}, {Steinmetz}, {Stewart},
  {Str{\"u}der}, {Sunyaev}, {Tenzer}, {Tiedemann}, {Tr{\"u}mper}, {Voron},
  {Weber}, {Wilms}, \& {Yaroshenko}}]{predehl2021}
{Predehl}, P., {Andritschke}, R., {Arefiev}, V., {et~al.} 2021, \aap, 647, A1,
  \dodoi{10.1051/0004-6361/202039313}

\bibitem[{{Rines} {et~al.}(2013){Rines}, {Geller}, {Diaferio}, \&
  {Kurtz}}]{rin2013}
{Rines}, K., {Geller}, M.~J., {Diaferio}, A., \& {Kurtz}, M.~J. 2013, \apj,
  767, 15, \dodoi{10.1088/0004-637X/767/1/15}

\bibitem[{{Rines} {et~al.}(2016){Rines}, {Geller}, {Diaferio}, \&
  {Hwang}}]{rin2016}
{Rines}, K.~J., {Geller}, M.~J., {Diaferio}, A., \& {Hwang}, H.~S. 2016, \apj,
  819, 63, \dodoi{10.3847/0004-637X/819/1/63}

\bibitem[{{Rudick} {et~al.}(2011){Rudick}, {Mihos}, \& {McBride}}]{rud11}
{Rudick}, C.~S., {Mihos}, J.~C., \& {McBride}, C.~K. 2011, \apj, 732, 48,
  \dodoi{10.1088/0004-637X/732/1/48}

\bibitem[{{Sampaio-Santos} {et~al.}(2020){Sampaio-Santos}, {Zhang}, {Ogando},
  {Shin}, {Golden-Marx}, {Yanny}, {Herner}, {Hilton}, {Choi}, {Gatti}, {Gruen},
  {Hoyle}, {Rau}, {De Vicente}, {Zuntz}, {Abbott}, {Aguena}, {Allam}, {Annis},
  {Avila}, {Bertin}, {Brooks}, {Burke}, {Carrasco Kind}, {Carretero}, {Chang},
  {Costanzi}, {da Costa}, {Diehl}, {Doel}, {Everett}, {Evrard}, {Flaugher},
  {Fosalba}, {Frieman}, {Garcia-Bellido}, {Gaztanaga}, {Gerdes}, {Gruendl},
  {Gschwend}, {Gutierrez}, {Hinton}, {Hollowood}, {Honscheid}, {James},
  {Jarvis}, {Jeltema}, {Kuehn}, {Kuropatkin}, {Lahav}, {Maia}, {March},
  {Marshall}, {Miquel}, {Palmese}, {Paz-Chinchon}, {Plazas}, {Sanchez},
  {Santiago}, {Scarpine}, {Schubnell}, {Smith}, {Suchyta}, {Tarle}, {Tucker},
  {Varga}, \& {Wechsler}}]{sam2020}
{Sampaio-Santos}, H., {Zhang}, Y., {Ogando}, R.~L.~C., {et~al.} 2020, MNRAS,
  submitted (arXiv:2005.12275), arXiv:2005.12275.
\newblock \doarXiv{2005.12275}

\bibitem[{{Sanderson} {et~al.}(2009){Sanderson}, {Edge}, \& {Smith}}]{san09}
{Sanderson}, A. J.~R., {Edge}, A.~C., \& {Smith}, G.~P. 2009, \mnras, 398,
  1698, \dodoi{10.1111/j.1365-2966.2009.15214.x}

\bibitem[{{Shaw} {et~al.}(2006){Shaw}, {Weller}, {Ostriker}, \& {Bode}}]{sha06}
{Shaw}, L.~D., {Weller}, J., {Ostriker}, J.~P., \& {Bode}, P. 2006, \apj, 646,
  815, \dodoi{10.1086/505016}

\bibitem[{{Sohn} {et~al.}(2018){Sohn}, {Chon}, {B{\"o}hringer}, {Geller},
  {Diaferio}, {Hwang}, {Utsumi}, \& {Rines}}]{sohn2018}
{Sohn}, J., {Chon}, G., {B{\"o}hringer}, H., {et~al.} 2018, \apj, 855, 100,
  \dodoi{10.3847/1538-4357/aaac7a}

\bibitem[{{Sohn} {et~al.}(2020){Sohn}, {Fabricant}, {Geller}, {Hwang}, \&
  {Diaferio}}]{sohn2020}
{Sohn}, J., {Fabricant}, D.~G., {Geller}, M.~J., {Hwang}, H.~S., \& {Diaferio},
  A. 2020, \apj, 902, 17, \dodoi{10.3847/1538-4357/abb23b}

\bibitem[{{Song} {et~al.}(2017){Song}, {Hwang}, {Park}, \& {Tamura}}]{son19}
{Song}, H., {Hwang}, H.~S., {Park}, C., \& {Tamura}, T. 2017, \apj, 842, 88,
  \dodoi{10.3847/1538-4357/aa72dc}

\bibitem[{{Springel} {et~al.}(2001){Springel}, {White}, {Tormen}, \&
  {Kauffmann}}]{spr01}
{Springel}, V., {White}, S. D.~M., {Tormen}, G., \& {Kauffmann}, G. 2001,
  \mnras, 328, 726, \dodoi{10.1046/j.1365-8711.2001.04912.x}

\bibitem[{{Springel} {et~al.}(2018){Springel}, {Pakmor}, {Pillepich},
  {Weinberger}, {Nelson}, {Hernquist}, {Vogelsberger}, {Genel}, {Torrey},
  {Marinacci}, \& {Naiman}}]{spr2018}
{Springel}, V., {Pakmor}, R., {Pillepich}, A., {et~al.} 2018, \mnras, 475, 676,
  \dodoi{10.1093/mnras/stx3304}

\bibitem[{{Umetsu}(2020)}]{ume20}
{Umetsu}, K. 2020, \aapr, 28, 7, \dodoi{10.1007/s00159-020-00129-w}

\bibitem[{{Umetsu} \& {Diemer}(2017)}]{ume17}
{Umetsu}, K., \& {Diemer}, B. 2017, \apj, 836, 231,
  \dodoi{10.3847/1538-4357/aa5c90}

\bibitem[{{Vikhlinin} {et~al.}(2009){Vikhlinin}, {Kravtsov}, {Burenin},
  {Ebeling}, {Forman}, {Hornstrup}, {Jones}, {Murray}, {Nagai}, {Quintana}, \&
  {Voevodkin}}]{vik2009}
{Vikhlinin}, A., {Kravtsov}, A.~V., {Burenin}, R.~A., {et~al.} 2009, \apj, 692,
  1060, \dodoi{10.1088/0004-637X/692/2/1060}

\bibitem[{{Vogelsberger} {et~al.}(2014){Vogelsberger}, {Genel}, {Springel},
  {Torrey}, {Sijacki}, {Xu}, {Snyder}, {Bird}, {Nelson}, \&
  {Hernquist}}]{vog14}
{Vogelsberger}, M., {Genel}, S., {Springel}, V., {et~al.} 2014, \nat, 509, 177,
  \dodoi{10.1038/nature13316}

\bibitem[{{Wang} {et~al.}(2009){Wang}, {Mo}, {Jing}, {Guo}, {van den Bosch}, \&
  {Yang}}]{wang2009}
{Wang}, H., {Mo}, H.~J., {Jing}, Y.~P., {et~al.} 2009, \mnras, 394, 398,
  \dodoi{10.1111/j.1365-2966.2008.14301.x}

\bibitem[{{West} {et~al.}(1995){West}, {Cote}, {Jones}, {Forman}, \&
  {Marzke}}]{west1995}
{West}, M.~J., {Cote}, P., {Jones}, C., {Forman}, W., \& {Marzke}, R.~O. 1995,
  \apjl, 453, L77, \dodoi{10.1086/309748}

\bibitem[{{Zahid} {et~al.}(2016){Zahid}, {Geller}, {Fabricant}, \&
  {Hwang}}]{zah2016}
{Zahid}, H.~J., {Geller}, M.~J., {Fabricant}, D.~G., \& {Hwang}, H.~S. 2016,
  \apj, 832, 203, \dodoi{10.3847/0004-637X/832/2/203}

\bibitem[{{Zhao}(1996)}]{zha96}
{Zhao}, H. 1996, \mnras, 278, 488, \dodoi{10.1093/mnras/278.2.488}

\bibitem[{{Zinger} {et~al.}(2018){Zinger}, {Dekel}, {Birnboim}, {Nagai}, {Lau},
  \& {Kravtsov}}]{zin18}
{Zinger}, E., {Dekel}, A., {Birnboim}, Y., {et~al.} 2018, \mnras, 476, 56,
  \dodoi{10.1093/mnras/sty136}

\end{thebibliography}
\bibliographystyle{aasjournal}

\end{document}